# Self-mixing in microtubule-kinesin active fluid from nonuniform to uniform distribution of activity

Teagan E Bate,[1] Megan E Varney,[2] Ezra H Taylor,[1] Joshua H Dickie,[1] Chih-Che Chueh,[3] Michael M Norton,[4] and Kun-Ta Wu[1,5,6,*]

[1]Department of Physics, Worcester Polytechnic Institute, Worcester, Massachusetts 01609, USA
[2]Department of Physics, New York University, New York, New York 10003, USA
[3]Department of Aeronautics and Astronautics, National Cheng Kung University, Tainan 701, Taiwan
[4]School of Physics and Astronomy, Rochester Institute of Technology, Rochester, New York 14623, USA
[5]Department of Mechanical Engineering, Worcester Polytechnic Institute, Worcester, Massachusetts 01609, USA
[6]The Martin Fisher School of Physics, Brandeis University, Waltham, Massachusetts 02454, USA
*Corresponding: kwu@wpi.edu

## Abstract

Active fluids have applications in micromixing, but little is known about the mixing kinematics of systems with spatiotemporally-varying activity. To investigate, UV-activated caged ATP was used to activate controlled regions of microtubule-kinesin active fluid and the mixing process was observed with fluorescent tracers and molecular dyes. At low Péclet numbers (diffusive transport), the active-inactive interface progressed toward the inactive area in a diffusion-like manner that was described by a simple model combining diffusion with Michaelis-Menten kinetics. At high Péclet numbers (convective transport), the active-inactive interface progressed in a superdiffusion-like manner that was qualitatively captured by an active-fluid hydrodynamic model coupled to ATP transport. Results showed that active fluid mixing involves complex coupling between distribution of active stress and active transport of ATP and reduces mixing time for suspended components with decreased impact of initial component distribution. This work will inform application of active fluids to promote micromixing in microfluidic devices.

## Introduction

Miniaturization enhances production efficiency in chemical engineering, biological engineering, and pharmaceutical manufacturing.[1] For example, microreactors—millimeter-scale devices with channels to mix chemicals and induce chemical reactions—are used to synthesize materials,[2] test enzymes,[3] and analyze protein conformations.[4] These devices require mixing to homogenize reactants, which is challenging because fluid dynamics at the micron scale are dominated by laminar flow. Mixing at a macroscopic scale is achieved by turbulence-induced advection repeatedly stretching and folding components until a uniform state is reached,[5] but at a microscopic scale, turbulence is inhibited (Reynolds number $\ll$ 1) and mixing is dominated by molecular diffusion, which is slow and difficult to control. Approaches such as serpentine design[6] and vibrating bubbles[7] have been developed to enhance micromixing, but these are driven by external energy sources and thus require external components that limit miniaturization.[1,8]

Active fluids—fluids with microscopic constituents that consume local fuel to generate movement[9-18]—have the potential to enhance mixing at the micron scale. Active fluids self-organize into chaotic turbulence-like flows[19-23] that promote micromixing by repeatedly stretching and folding fluid, even at low Reynolds numbers.[24] Prior work on active mixing has focused on active systems with uniform activity distribution.[24-26] However, mixing processes often start from a state of nonuniformity. Nonuniform distributions of activity in active fluid systems can cause complex dynamics.[27-34] Spatiotemporal patterns of activity that are prescribed from an external source[31,32] or emerge as an additional dynamical variable that coevolves with system[27-29] have been studied. However, the effect of nonuniform distributions of activity on mixing has not been elucidated.

Here, we studied the mixing dynamics of a microtubule-kinesin suspension whose activity is governed by the transport of ATP, the system's energy source. We controlled the initial distribution of ATP by using caged ATP that can only fuel the fluid after exposure to ultraviolet (UV) light. This allowed us to repeatedly observe the transient dynamics that carry the system from heterogeneous activity to homogeneous activity. We explored mixing dynamics ranging from diffusion-dominated to convection-dominated by varying the ATP concentration, composition of kinesin motors, flow cell geometry, and initial distribution of ATP. We contextualized the results with models at two levels of complexity. A simple model captured the mixing dynamics in a diffusion-limited regime, whereas a more complex model that included active-fluid hydrodynamics reproduced aspects of observed enhanced transport and activity-dependent progression of the active-inactive interface.

## Results

**Active and inactive fluids self-mix into a homogeneous active fluid.** For the experiments presented herein, we selected a 3D microtubule-kinesin active fluid because it enhances micromixing,[14,24] has tunable activity,[35-40] and has established models describing its flow behaviors.[41-44] In microtubule-kinesin active fluid, microtubules self-assemble by depletion into bundles that extend spontaneously, driving chaotic vortical flows. The extension is driven by kinesin motor dimers that hydrolyze ATP to "walk" along pairs of antiparallel microtubules and force them in opposite directions (Fig. 1a).[14] We augmented the microtubule-kinesin system with UV light-activated chemistry that allowed us to create distinct patterns of activity. In this light-activated system, the ATP is "caged"—its terminal phosphate is esterified with a blocking group (Fig. 1b)—such that it cannot be hydrolyzed by kinesin motors until the blocking group is removed by exposure to UV light.[45,46] In this system, the activation of the fluid is irreversible. After the fluid is activated, the action of the kinesin motors causes the microtubule network to become a 3D self-rearranging isotropic active gel consisting of extensile microtubule bundles that buckle and anneal repeatedly until the ATP is exhausted.[14] To quantify the evolution of the activity distribution, we suspended fluorescent tracers in the solvent and monitored the tracer motion to extract the speed distribution of active fluid flows (Fig. 1d). To observe the structure of the active suspension, we labelled microtubules with Alexa 647 (Fig. 1c).

When the fluid was in its inactive state, before it had been activated by UV light, the kinesin motor dimers were bound to microtubules, creating a quiescent crosslinked microtubule network that behaved like an elastic gel (Fig. 1c, top panel). The inactive gel was essentially isotropic, but after the fluid was loaded into a rectangular flow cell ($20 \times 4 \times 0.1$ mm$^3$) we observed some alignment of the bundles near the boundary (Supplementary Fig. 1).[47] To create an active-inactive interface, we used a mask to apply UV light to one side of the sample, which released the ATP and activated the fluid on that side only (Fig. 1c, second panel; Supplementary Video 1). The spatial pattern of activity evolved from a sharp interface to become increasingly diffuse as the initially active region invaded the inactive region (Fig. 1d, second and third panels). We quantified this evolution of activity distribution with the normalized speed profile (Fig. 1e), which shows how the interface between regions widened and shifted as the active and inactive parts of the microtubule system blended over a period of hours.

**When active fluid mixes with inactive fluid, are the mixing dynamics governed by a superdiffusion-like process?** Active fluids enhance the motion of suspended tracers from diffusive (having a mean squared displacement [MSD] proportional to time lapse: MSD $\sim \Delta t^a$ with $a = 1$) to superdiffusive ($a > 1$).[14,26,35] The progression of the active-inactive interface can also be described as diffusion-like or superdiffusion-like as follows: Suppose the displacement of the active-inactive interface is $\Delta x$ and the squared interface displacement increases with time as $\Delta x^2 \sim 2P_I \, t^\gamma$ with the interface progression coefficient $P_I$ and the

interface progression exponent $\gamma$. If $\gamma = 1$, the progression of the active-inactive interface is defined as diffusion-like; if $\gamma > 1$, the progression of the active-inactive interface is defined as superdiffusion-like.

Because active fluids enhance microscale transport, we hypothesized that the active-inactive interface would progress in a superdiffusion-like manner ($\gamma > 1$). To test this hypothesis, we quantified the displacement of the active-inactive fluid interface as a function of time (Fig. 2a inset) and found that motion of the interface decelerated as the active fluid mixed with the inactive fluid such that the squared interface displacement progressed as $\Delta x^2 \sim t^\gamma$ with an interface progression exponent $\gamma \approx 1$ (Fig. 2a). We repeated the $\gamma$ measurement for caged ATP concentrations from 0.5 to 8 mM (0.5 mM is enough to maximize the flow speed of active fluid[48]) and consistently found that $\gamma \approx 1$ across this range (Fig. 2b). These results invalidated our hypothesis and suggested that the progression of active-inactive interface is diffusion-like. Notably, while the diffusive time-scaling $\gamma$ remained consistent, the prefactor $P_I$ exhibited a monotonic but nonlinear dependence on caged ATP concentration (Fig. 2c).

**Modeling results showed that the mixing process of active and inactive fluids is governed by diffusion-like processes of ATP at the active-inactive interface.** In the experiments, the interface progression coefficient $\gamma \approx 1$, which suggested that diffusion dominated the dynamics of the active-inactive interface. To contextualize our observations, we constructed a minimal model that combines diffusion of ATP with a previously measured relation between ATP concentration and local fluid velocity.[48] Herein, we modeled ATP's dispersion using Fick's law of diffusion:

$$\frac{\partial C(\mathbf{r}, t)}{\partial t} = D\nabla^2 C(\mathbf{r}, t), \qquad 1$$

where $C(\mathbf{r}, t)$ represents the spatial distribution of ATP concentrations at time $t$ and $D$ is the diffusion coefficient of ATP in active fluid. We chose $D = 140$ μm²/s, which is one-fifth the diffusion coefficient of ATP in water,[49] because the crosslinked microtubule network makes the fluid more viscous than water[50] and our measurement on diffusion coefficient of suspended fluorescein was one-fifth of its reported value in aqueous solution (Supplementary Discussion 1). To simplify the modeling, we considered a 1D active fluid system confined in a segment, $x = 0 - L$, where $L = 20$ mm is the segment length, and applied no-flux boundary conditions

$$\frac{\partial}{\partial x} C(x = 0, t) = \frac{\partial}{\partial x} C(x = L, t) = 0. \qquad 2$$

To mimic the UV-activation process (Fig. 1c), we initiated the ATP concentration with a step function

$$C(x, t = 0) = \frac{C_0}{2} \text{erfc}\left(\frac{x - x_0}{2\sqrt{\epsilon D}}\right), \qquad 3$$

where $C_0 = 0.5-8$ mM, the initial ATP concentrations in the activated region; erfc is the complementary error function; and $x_0 = 10$ mm is the initial position of the active-inactive interface. We chose $\epsilon = 0.001$ to generate a sharp concentration transition at the interface. We numerically solved Eqs. 1-3 to determine the spatial and temporal distribution of ATP concentrations (Fig. 3a; Supplementary Video 2). To relate the evolving ATP distribution to local flow speed, we leverage previous experimental results[48] that found the average velocity in bulk samples follows Michaelis-Menten kinetics

$$\bar{v}(x, t) = \bar{v}_m \left[\frac{C(x, t)}{C(x, t) + K}\right], \qquad 4$$

where $\bar{v}$ is the mean speed of active fluid, $\bar{v}_m = 6.2$ μm/s is the saturated mean speed, and $K = 270$ μM is the ATP concentration that leads to half of the saturated mean speed, $\bar{v}_m/2$. (Model selection is described in Supplementary Discussion 2.) The mean speed distributions (Fig. 3b) showed that initially one side of the sample was activated (black) and then the sample evolved toward a more uniformly activated state (red). The squared interface displacement of the active-inactive interface increased linearly with time, $\Delta x^2 \sim t$ (Fig. 3c; see Supplementary Discussion 3 for derivation of $\Delta x^2 \propto t$), which matched our experimental observation of $\gamma \approx 1$ (Figs. 2a & 2b). Further, we compared the dependency of the interface progression coefficient $P_I$ (determined by fitting $\Delta x^2$ vs. $t$ data to $\Delta x^2 = 2P_I\,t$ with $P_I$ as the fitting parameter; Figs. 2c & 3c) on initial ATP concentrations between experiment and model and also found excellent agreement (differed by ~10%; Fig. 3d). Taken together, the agreement between simulation and experiment on the scaling of the dynamics (Figs. 2b & 3c) and dependency on initial ATP concentration (Fig. 3d) indicated that the dispersion of ATP is dominated by diffusion and that Michaelis-Menten kinetics are appropriate for a coarse-grained model to connect ATP concentration with local flow speed of active fluid,[48] without the need to introduce a more complex hydrodynamic model.[42,51]

**Superdiffusion-like progression of the active-inactive interface only emerges when active transport is dominated by convection.** The success of the diffusion-limited model suggests that the active transport in the active fluid systems studied above was dominated by diffusion. This inspired us to question whether the progression of active-inactive interface would become superdiffusion-like when the active transport becomes convection-like.[27,28] To answer this question, we varied experimental parameters to explore a wider range of fluid flow speeds. To achieve lower flow speeds, we altered the composition of motor proteins by replacing a fraction of the processive motors (K401), which exert force on microtubules continuously, with nonprocessive motors (K365) that detach after each force application. The reduced number of processive motors has the net effect of driving the extensile motion of microtubules more slowly (Fig. 4a inset left).[36,48] To achieve higher flow speeds, we increased the height of the sample container to decrease hydrodynamic drag (Fig. 4a inset right).[48,52] Throughout these experiments, we kept the caged ATP concentration constant (5 mM).

As in the previous experiments, we analyzed the spatiotemporal progression of activity to find the interface progression exponent $\gamma$ as a function of the average flow speed in the bulk of the initially activated area, $\bar{v}_{ab}$ (Fig. 4a). Because changing channel geometry alters the characteristic size of vortices in active fluids,[39] we unified our datasets by plotting $\gamma$ as a function of the Péclet number, $Pe$ (Fig. 4b), defined as $Pe \equiv \bar{v}_{ab}l_c/D$ where $l_c$ is the correlation length of flow velocity (see Supplementary Discussion 4) and $D = 140$ μm²/s is our estimate of ATP diffusion in the system (see Supplementary Discussion 1).[53,54] The Péclet number is a dimensionless quantity representing the ratio of convective transport rate to diffusive transport rate. A larger Péclet number (typically of order 10 or above) indicates convection-dominated active transport, and a smaller Péclet number (typically of order 1 or below) indicates diffusion-dominated active transport. Our data showed that for $Pe \lesssim 3$, the interface progression exponent remained $\gamma \approx 1$ (Fig. 4b), which corresponded to the regime captured by our model (Fig. 3). Then as $Pe$ increased to greater than 3, $\gamma$ grew monotonically (Fig. 4b). For the largest $Pe$ explored in our experiments ($Pe \approx 16$), $\gamma$ reached ~1.7, which indicated that convective processes were beginning to emerge and dominate the active transport. Overall, our data suggested that as the active transport transitioned from diffusion-dominated ($Pe \lesssim 3$) to convection-dominated ($Pe \gtrsim 3$) regimes, the progression of active-inactive interfaces transitioned from diffusion-like ($\gamma \approx 1$) to superdiffusion-like ($\gamma > 1$).

**Active fluid flows reduce the mixing time of UV-activated fluorescent dyes.** To this point, we had characterized the mixing of active and inactive fluids by the progression of the interface between them; however, like milk blending into coffee, the mixing process often involves dispersion of suspended components. To characterize how suspended components disperse during the progression of the active-inactive interface, we designed another series of experiments with suspended components that were initially nonuniform. We doped inactive fluid with suspended UV-activated fluorescent dyes and exposed one side of the sample container to UV light, which simultaneously activated the fluid and the fluorescent dye. We found that in an inactive sample ($\bar{v}_{ab} = 0$), where dyes dispersed only by molecular diffusion, the dye barely dispersed, whereas in a sample where one side was activated ($\bar{v}_{ab} = 8.2$ μm/s), the dyes were transported by active fluid flows and almost completely dispersed through the sample in 4 hours (Fig. 5a). To quantify the dispersion rate, we adopted Saintillan and Shelley's method[25] to analyze the normalized multiscale norm of dye brightness as a function of time: $\hat{s}(t) \equiv |s(t)|/|s(0)|$, where

$$|s| \equiv \left[ \sum_{k} \frac{|s_k|^2}{\sqrt{1 + l^2 k^2}} \right]^{1/2},$$
5

$s_k$ is the Fourier coefficient at wave vector $\boldsymbol{k}$ in a Fourier expansion of the dye brightness and $l = 4.84$ μm is the pixel size of the micrographs. We found that the normalized multiscale norm decayed faster as $\bar{v}_{ab}$ increased from 0 to 8.2 μm/s (Fig. 5b). In light of reports that the norm decays exponentially,[25] we quantified the decay rate by fitting the first hour $\hat{s}(t)$ data to $\ln \hat{s} = -t/t_0$ with $t_0$ (mixing time) as the fitting parameter (Fig. 5b inset) and found that the mixing time decreased with flow speed of active fluid (Fig. 5c inset). When the fluid was inactive ($\bar{v}_{ab} = 0$), dye dispersion was dominated by molecular diffusion and the mixing time was 24 hours; slightly activating the fluid ($\bar{v}_{ab} = 2$ μm/s) reduced the mixing time to 8 hours, which demonstrated that active fluid flows enhanced the mixing process of suspended components.[26]

To reveal how the mechanism of active transport (i.e., diffusion-dominated or convection-dominated) altered the mixing time, we analyzed the mixing time as a function of the Péclet number and found that the mixing time monotonically decreased as the active transport became more convection-like (Fig. 5c). Notably, there was no discernible transition in mixing time as the active transport transitioned from diffusion-dominated to convection-dominated, although there was a transition in the progression of active-inactive interfaces (Fig. 4b). This dependence of mixing time on the Péclet number in active-inactive fluid systems was similar to that in an activity-uniform active fluid system (Supplementary Discussion 5 and Supplementary Fig. 6b), which showed that Péclet number was the controlling parameter for mixing time of suspended components in active fluid systems, regardless of the distribution of activity.

**A continuous active fluid model captures the mixing of nonuniform active fluid systems.** Our experimental data showed that as the active transport became more convection-like, the active-inactive interface progression transitioned from diffusion-like to superdiffusion-like (Fig. 4b) and the mixing time of suspended components decreased monotonically (Fig. 5c). To determine whether this complex mixing process could be modeled with an existing active fluid model, we adopted Varghese *et al.*'s model[51] because it successfully describes the transition from coherent to chaotic flow in 3D microtubule-kinesin active fluid systems.[52] The model describes microtubules as self-elongating rods whose nematic order, $\boldsymbol{Q}$, is subject to spontaneous decay due to the rods' rotational molecular diffusion and reorientation by solvent flow. Thus, the dimensionless kinetic equation for $\boldsymbol{Q}$ can be written as:

$$\partial_{t^*} \boldsymbol{Q} + \boldsymbol{u} \cdot \nabla_* \boldsymbol{Q} + \boldsymbol{Q} \cdot \boldsymbol{\Omega}^* - \boldsymbol{\Omega}^* \cdot \boldsymbol{Q} = -\boldsymbol{Q} + \nabla_*^2 \boldsymbol{Q} + \lambda \left[ \frac{2}{d} \boldsymbol{E}^* + \boldsymbol{Q} \cdot \boldsymbol{E}^* + \boldsymbol{E}^* \cdot \boldsymbol{Q} - \frac{2}{d} \mathrm{Tr}(\boldsymbol{Q} \cdot \boldsymbol{E}^*) \boldsymbol{I} \right],$$
6

where $t^*$ is the dimensionless time, $\nabla_*$ is the dimensionless spatial gradient operator, $\nabla_*^2$ is the dimensionless Laplacian operator, $\boldsymbol{\Omega}^* \equiv [(\nabla_* \boldsymbol{u})^T - \nabla_* \boldsymbol{u}]/2$ is the dimensionless vorticity tensor, $\boldsymbol{E}^* \equiv$ $[(\nabla_* \boldsymbol{u})^T + \nabla_* \boldsymbol{u}]/2$ is the dimensionless strain rate tensor, $\lambda = 1$ is the flow alignment coefficient, and $d$ is the system dimensionality. The dimensionless flow field $\boldsymbol{u}$ is governed by the Stokes equation

$$\nabla_*^2 \boldsymbol{u} - \nabla_* p^* - \nabla_* \cdot \boldsymbol{\sigma}_a = 0 \qquad\qquad 7$$

and incompressibility constraint ($\nabla_* \cdot \boldsymbol{u} = 0$), where $p^*$ is the dimensionless pressure and $\boldsymbol{\sigma}_a \equiv \alpha^* \boldsymbol{Q}$ is the dimensionless active stress exerted by self-elongating rods with a dimensionless activity coefficient $\alpha^*$.[55] Because the activity coefficient increases with ATP concentration,[56] we selected an $\alpha$-ATP relation[57]

$$\alpha^* = \alpha_0^* \frac{C}{C + K} \, , \qquad\qquad 8$$

where $\alpha_0^*$ is the dimensionless activity level, $C$ is the ATP concentration, and $K = 270$ μM.[48] We selected this relation because it captures the dynamics of microtubule bundle extension and kinesin kinetics (Michaelis-Menten), which play critical roles in the activity of microtubule-kinesin active fluid systems.[58,59] Finally, given that ATP diffused as a result of thermal fluctuation as well as flowed with the active fluid, we modeled ATP dispersion with a convection-diffusion equation:

$$\partial_{t^*} C = D^* \nabla_*^2 C - \boldsymbol{u} \cdot \nabla_* C, \qquad\qquad 9$$

where $D^*$ is the dimensionless ATP molecular diffusion coefficient. To simplify modeling, we considered a 2D active fluid system ($d = 2$)[51] confined in a $112 \times 22$ rectangular boundary with no-slip boundary condition for flows ($\boldsymbol{u} = \boldsymbol{0}$) and no-flux boundary condition for rods ($\boldsymbol{n} \cdot \nabla_* \boldsymbol{Q} = \boldsymbol{0}$, where $n$ represents a unit vector normal to boundaries). To solve the equations for $\boldsymbol{Q}$, $\boldsymbol{u}$, and $C$ (Eqs. 6, 7, & 9), we determined the initial conditions as quiescent solvent ($\boldsymbol{u} = \boldsymbol{0}$) under uniform pressure ($p^* = 0$) with the rods in an isotropic state [$Q_{xx} = -Q_{yy} = 2.5 \times 10^{-4} \, rn(\boldsymbol{r})$ and $Q_{xy} = Q_{yx} = 5 \times 10^{-4} \mathrm{rn}(\boldsymbol{r})$, where $\mathrm{rn}(\boldsymbol{r})$ is a spatial uniform random number between $-1$ and $+1$] and 5 mM of ATP distributed on only one side of the system. Then we evolved the fluid flows and ATP distributions for 200 units of dimensionless time ($t^* = 0$–200) with the finite element method.[60]

Our modeling results (Supplementary Video 4) showed that in an inactive system ($\alpha_0^* = 0$; $D^* = 16$; Fig. 6a, left column), ATP dispersed only by molecular diffusion, but when one side of the sample was activated ($\alpha_0^* = 25$; $D^* = 16$; middle column), the system developed chaotic turbulence-like mixing flows that actively transported the ATP toward the inactive region. In a third simulation where the ATP molecular diffusion rate was increased ($\alpha_0^* = 25$; $D^* = 64$; right column), the mixing process sped up. These simulation results showed that the process of ATP dispersion was controlled by both molecular diffusion of ATP and active fluid-induced convection.

To quantify the efficacy of ATP mixing by active fluid, we analyzed the normalized multiscale norm of ATP concentrations as a function of time[25] for $\alpha_0^* = 0$–25 and $D^* = 1$–128 (Eq. 5 with $l = 1$). We found that the norms decayed exponentially with time: $\hat{s} \sim \exp(-t^*/t_0^*)$, where $t_0^*$ is the dimensionless mixing time (Fig. 6b), which was consistent with results reported by Saintillan and Shelley.[25] We analyzed mixing time as a function of dimensionless activity level, $\alpha_0^*$, for each dimensionless molecular diffusion coefficient $D^*$ (Fig. 6c) and found that when ATP diffused slowly ($D^* \lesssim 16$; blue to light green curves), mixing time decreased with increasing activity level or faster active transport (Fig. 6b inset), which was consistent with our experimental observation (Fig. 5c). Our simulation also showed that as ATP diffused sufficiently fast ($D^* \gtrsim 32$; olive and red curves), the mixing time was nearly independent of activity level. Overall, increasing both the molecular diffusion coefficient and the activity level decreased mixing time. Thus, our simulation showed that both molecular diffusion (represented as $D^*$) and active fluid-induced convection (related to $\alpha_0^*$) played important roles to disperse and homogenize the suspended components; which mechanism dominated the dispersion depended upon the competition between these two mechanisms.

To demonstrate how the competition of these two mechanisms affected the progression of active-inactive interfaces, we analyzed the interface progression exponent $\gamma$ as a function of $\alpha_0^*$ for various diffusion coefficients $D^*$ (Fig. 6c inset) and found that when the diffusion mechanism was relatively weak ($D^* = 2$; dark blue curve), the convection mechanism dominated the interface progression, leading it to progress in a superdiffusion-like, or more precisely, ballistic-like manner ($\gamma \approx 2$). Contrarily, as the diffusion mechanism became relatively strong ($D^* = 8$; light green curve), diffusion mechanisms dominated the interface progression, leading it to progress in a diffusion-like manner ($\gamma \approx 1$). Interestingly, we found that in an intermediate strength of diffusion mechanism ($D^* = 4$; dark green curve), increasing activity level $\alpha_0^*$ transitioned the interface progression from diffusion-like to superdiffusion-like, which is consistent with our experimental observation (Fig. 4). Overall, our active-fluid hydrodynamic model qualitatively captures the mixing dynamics of active and inactive fluid systems in terms of active-inactive interface progression and dispersion of suspended components.

**Fluid activity enhances mixing of suspended components and reduces impact of nonuniformity of component distribution on mixing time.** Up until this point we had only explored one configuration of nonuniform active fluid systems: an activated bulk on one side of a channel adjacent to an inactive bulk on the other side. To explore how other spatial configurations of activity affect mixing, we used a checkerboard pattern of UV light to split the activated region into cells. As in previous experiments, 50% of the total fluid was activated. Fluid activated in a checkerboard pattern evolved to a homogeneous state more quickly than fluid that was activated on one side only (1 hour vs. 10 hours; Fig. 7a). UV-activated fluorescent dyes showed that the mixing time decreased as the grid size decreased from 3 mm to 1 mm (Fig. 7b).

To elucidate how checkerboard mixing driven by active fluid differed from that driven by molecular diffusion alone, we applied our established active-fluid hydrodynamic model for both active ($\alpha^* = 25$) and inactive ($\alpha^* = 0$) fluid systems (Fig. 8a; Supplementary Video 6). As expected, we found that the mixing time increased monotonically with grid size for both active and inactive fluid systems (Fig. 8b), with the active fluid system (red curve) having a shorter mixing time than the inactive fluid system (black curve). Interestingly, we found that when the grid size was sufficiently small ($a \lesssim 5$), the active and inactive fluids had the same simulated mixing time. We also found that as the grid size increased from 5 to 22, the mixing time of the inactive fluid increased more than the mixing time of the active fluid (40× vs. 3×).

### Discussion

The self-mixing process of microtubule-kinesin active fluid with nonuniform activity was driven by active transport at the active-inactive interface. We estimated the contributions of diffusive and convective transport using the Péclet number, $Pe$. We found that when the active transport was dominated by the diffusion mechanism ($Pe \lesssim 3$), the active-inactive interface progressed in a diffusion-like manner ($\gamma \approx 1$; Fig. 2). These dynamics were quantitatively captured by a Fick's law-based model that quasi-statically related local activity to the local concentration of ATP by using a previously measured ATP-velocity relation (Fig. 3).[48]

As we raised the Péclet number ($Pe \gtrsim 3$) by increasing both the local fluid velocity and mixing length scale, we found that the active-inactive interface concomitantly progressed in a more superdiffusion-like manner ($\gamma > 1$; Fig. 4). We observed experimentally that increasing the Péclet number decreased the mixing time of suspended fluorescent dyes (Fig. 5c), which demonstrated that more convective flow mixed the suspended components faster. These results, along with the progression of the active-inactive interfaces, were qualitatively captured by an active-fluid hydrodynamic model (Fig. 6c) that coupled active stress-induced fluid flow and transport of ATP molecules (Eqs. 6-9).

Interestingly, while our hydrodynamic model predicted interface progression exponent $\gamma = 2$ for high activity levels (Fig. 6c inset), in our experiments $\gamma$ appeared to plateau at $\gamma \approx 1.7$ (Fig. 4b). Our model may have overestimated $\gamma$ because the microtubule network in the inactive portion of the sample is crosslinked by immobile kinesin motor dimers that cause it to behave like an elastic gel. When the ATP molecules arrive at the active-inactive interface, they fuel the motor dimers, which fluidizes the network. Crucially, this fluidizing/melting process takes time to develop.[47,50] Thus, for the interface to progress, not only does ATP need to be transported to the inactive fluid region, but the ATP-fueled motors also needs time to melt the gel-like microtubule network into a fluid. Such melting dynamics could delay the progression of the active-inactive interface and lower $\gamma$. In the simulation, the melting dynamics were absent; the network melted almost instantly as soon as ATP arrived at the inactive fluid, and $\gamma$ would only depend on active transport of ATP. Our additional studies (Supplementary Discussion 6) support the idea of a network melting mechanism by showing that the progression of the active-inactive interface fell behind the progression of ATP molecules (Supplementary Fig. 7e), whereas in the simulation the fronts of both coincided (Supplementary Fig. 8c). Future research to elucidate the network melting dynamics could involve monitoring dyes, tracers, and microtubules simultaneously to reveal the correlations among ATP dispersion, active fluid flows, and microtubule network structure (melting). The process could be modeled with the active-fluid hydrodynamic model used herein, modified to include ATP-dependent rheological constants and additional relevant dynamic processes to represent the melting process of the gel-like network at the interface.

We also found that the distribution of activity had a significant effect on mixing time. Systems consisting of more, smaller active areas (checkerboard pattern; high uniformity [Fig. 7]) evolved to a homogeneous state faster than systems with the same total active area distributed as one piece (one side active; low uniformity [Fig. 1c]). This is likely because the smaller grid size increased the active-inactive interface area, which allowed the active fluid to interact with inactive fluid more efficiently. Interestingly, our active-fluid hydrodynamic model showed that when the grid size was sufficiently small, the mixing times of active and inactive fluids were indistinguishable (Fig. 8b). This may be because the active fluid needs time to "warm up" from an initial quiescent state before reaching its steady activity state.[47] In experiments, the system had a warm-up time that may have been caused by network melting (Supplementary Discussion 6). Although a network melting mechanism was not included in the model, the simulated active fluid flow took dimensionless time to rise because the onset of the flows was triggered by the initial activity-driven instability in extensile $Q$ field, which took finite dimensionless time to develop (~1 dimensionless time in this case; Supplementary Video 6).[25,47] Thus, in cases where the grid size was sufficiently small, the model showed molecular diffusion completing the mixing before emergence of active fluid flows. We also found that mixing time in an active fluid system was less sensitive to initial distribution of activity than that in an inactive system (Fig. 8b), which suggests that introducing active fluid to a microfluidic system could drastically reduce the impact of the initial condition on mixing efficacy.

This study has limitations. Observations on the mixing of active microtubule-kinesin fluid and inactive microtubule-kinesin fluid may not be generalizable to cases in which active fluid mixes with other types of fluid. Also, we did not characterize the degree of chaos in the system, such as by measuring Lyapunov exponents and topological entropies.[24] Future research could track tracers in 3D and measure how these quantities change in the 3D isotropic active microtubule network at different strengths of active transport (i.e., Lyapunov exponent vs. Péclet number and topological entropy vs. Péclet number).

Another limitation of this study was that our results for interface progression transitioning from diffusion-like to superdiffusion-like (Fig. 4) were based on large length-scale data that we analyzed considering the interface as one piece with a specific position coordinate (Fig. 2a). However, the interface is the region where ATP concentration decays from saturation ($> K$; see Eq. 4) to 0, and according to previous studies[14,35] tracer motion in this region should transition from superdiffusion-like to diffusion-like behaviors. Directly measuring the mean squared displacement of tracers across the active-inactive interface would elucidate the transition of the interface progression behaviors on the microscopic scale. Such measurements were not practical in our system because the active-inactive interface changed position and width with time (Fig. 1e); tracers initially in the diffusive zone could later be in the superdiffusive zone as the interface passed by, and it would be difficult to distinguish between the diffusive and superdiffusive zones. Future research to elucidate tracer behaviors at active-inactive interfaces could utilize fluid that is only active when it is exposed to light[31,32] to provide a stable activity gradient and thus obtain a reliable mean squared displacement of tracers at different parts of the interface.

Overall, this work demonstrated that mixing in nonuniform active fluid systems is fundamentally different from mixing in uniform active fluids. Mixing in nonuniform active fluid systems involves complex interplays among spatial distribution of ATP, active transport of ATP (which can be either diffusion-like or convection-like, depending on Péclet number), and a fluid-gel transition of the microtubule network at the interface.[47] This work paves the path to the design of microfluidic devices that use active fluid to promote or optimize the micromixing process[8] to enhance production efficiency in chemical and biological engineering and pharmaceutical development.[1] The results may also provide insight into intracellular mixing processes, because the cytoplasmic streaming that supports organelles within cells is powered by cytoskeletal filaments and motor proteins that function similarly to microtubule-kinesin active fluid.[61]

## Methods

**Polymerize microtubules.** Microtubules constitute the underlying network of microtubule-kinesin active fluid. Microtubules were polymerized from bovine brain $\alpha$- and $\beta$-tubulin dimers purified by three cycles of polymerization and depolymerization.[62,63] The microtubules (8 mg/mL) were then stabilized with 600 μM guanosine-5′[(α,β)-methyleno]triphosphate (GMPCPP, Jena Biosciences, NU-4056) and 1 mM dithiothreitol (DTT, Fisher Scientific, AC165680050) in microtubule buffer (80 mM PIPES, 2 mM MgCl₂, 1 mM ethylene glycol-bis(β-aminoethyl ether)-N,N,N',N'-tetraacetic acid, pH 6.8) and polymerized by a 30-minute incubation at 37 °C and a subsequent 6-hour annealing at room temperature before being snap frozen with liquid nitrogen and stored at −80 °C. The microtubules were then labeled with Alexa Fluor 647 (excitation: 650 nm; emission: 671 nm; Invitrogen, A-20006) and mixed with unlabeled microtubules at 3% labeling fraction during polymerization to image microtubules for non-fluorescein experiments (Figs. 1, 2, & 4). For fluorescein experiments, the microtubules were unlabeled (Figs. 5 & 7).

**Dimerize kinesin motor proteins.** Kinesin motor proteins power the extensile motion of sliding microtubule bundle pairs in active fluid by forming a dimer and walking on adjacent antiparallel microtubules to force them in opposite directions (Fig. 1a).[14,64] We expressed kinesin in the *Escherichia coli* derivative Rosetta 2 (DE3) pLysS cells (Novagen, 71403), which we transformed with DNA plasmids from *Drosophila melanogaster* kinesin (DMK) genes.[65] For most experiments in this paper, we used processive motors that include DMK's first 401 N-terminal DNA codons (K401).[66] To explore the effect of low mean speed of active fluid bulk on the interface progression exponent $\gamma$, we mixed in fractions of nonprocessive motors whose plasmid included DMK's first 365 codons (K365, Fig. 4 inset left).[36,48,67] The kinesin motors were tagged with 6 histidines enabling purification via immobilized metal ion affinity chromatography with gravity nickel columns (GE Healthcare, 11003399). To slide adjacent microtubule bundle pairs, kinesin motors needed to be dimerized, so the kinesin motors were tagged with a biotin carboxyl carrier protein at their N terminals, which allowed the kinesins to be bound with biotin molecules (Alfa Aesar, A14207).[14,62] To dimerize the kinesin, we mixed either 1.5 μM K401 processive motors or 5.4 μM K365 nonprocessive motors with 1.8 μM streptavidin (Invitrogen, S-888) and 120 μM DTT in microtubule buffer, incubated them for 30 minutes at 4 °C, and then snap froze them with liquid nitrogen and stored them at −80 °C.

**Prepare microtubule-kinesin active fluid with caged ATP and caged fluorescein.** To prepare the active fluid, we mixed 1.3 mg/mL microtubules with 120 nM kinesin motor dimers and 0.8% polyethylene glycol (Sigma 81300), which acted as a depleting agent to bundle microtubules (Fig. 1a).[14] Kinesin steps from the minus to the plus end of microtubules by hydrolyzing ATP and producing adenosine diphosphate.[59] To control the initial spatial distribution of ATP and thus the activity distribution of active fluid, we used 0.5 to 8 mM caged ATP (adenosine 5'-triphosphate, P3-(1-(4,5-dimethoxy-2-nitrophenyl)ethyl) ester, disodium salt and DMNPE-caged ATP, Fisher Scientific, A1049), which is ATP whose terminal phosphate is esterified with a blocking group rendering it nonhydrolyzable by kinesin motors unless exposed to 360-nm UV light. Exposure to UV light removes the blocking group (Fig. 1b) and allows the kinesin motors to hydrolyze the ATP into ADP and activate the active fluid.[45,46] The ATP hydrolyzation decreased ATP concentrations, which slowed down the kinesin stepping rate and thus decreased active fluid flow speed.[14,35,48,58,59] To maintain ATP concentrations so as to stabilize the activity level of the active fluid bulk over the course of our experiments, we included 2.8% v/v pyruvate kinase/lactate dehydrogenase (Sigma, P-0294), which converted ADP back to ATP.[14,68] To feed the pyruvate kinase enzyme, we added 26 mM phosphenol pyruvate (BeanTown Chemical, 129745). We imaged the active fluid samples with fluorescent microscopy for 1 to 16 hours, which could bleach the fluorescent dyes and thus decrease the image quality over the course of experiments. To reduce the photobleaching effect, we included 2 mM trolox (Sigma, 238813) and oxygen-scavenging enzymes consisting of 0.038 mg/mL catalase (Sigma, C40) and 0.22 mg/mL glucose oxidase (Sigma, G2133) and fed the enzymes with 3.3 mg/mL glucose (Sigma, G7528).[14] To stabilize proteins in our active fluid system, we added 5.5 mM DTT. To track the fluid flows, we doped the active fluid with 0.0016% v/v fluorescent tracer particles (Alexa 488-labeled [excitation: 499 nm; emission: 520 nm] 3-μm polystyrene microspheres, Polyscience, 18861). To test how active fluid could mix suspended components, we introduced 0.5 to 6 μM caged, UV-activated fluorescent dyes (fluorescein bis-(5-carboxymethoxy-2-nitrobenzyl) ether, dipotassium salt; CMNB-caged fluorescein, ThermoFisher Scientific, F7103), which are colorless and nonfluorescent until exposed to 360-nm UV light.[45] The dye concentration was chosen to maintain a sufficient signal-to-noise ratio while avoiding brightness saturation in micrographs. Upon UV exposure, the fluorescein was uncaged and thus became fluorescent and could be observed with fluorescent microscopy. Because the fluorescent spectrum of the fluorescein overlapped with our Alexa 488 tracers, for our experiments with caged fluorescein (Fig. 5) we replaced the tracers with Flash Red-labeled 2-μm polystyrene microspheres (Bangs Laboratories, FSFR005) and used unlabeled microtubules (0% labeling fraction) to prevent fluorescent interference from microtubules while imaging tracers.

**Prepare active-inactive fluid systems.** To prepare the active-inactive fluid system, we loaded the inactive microtubule-kinesin fluid with caged ATP to a polyacrylamide-coated glass flow cell (20 × 4 × 0.1 mm³) with Parafilm (Cole-Parmer, EW-06720-40) as a spacer sandwiched between a cover slip (VWR, 48366-227) and slide (VWR, 75799-268)[36] and sealed the channel with epoxy (Bob Smith Industries, BSI-201). Then we masked one side of the sample with a removable mask of opaque black tape (McMaster-Carr, 76455A21) attached to a transparent plastic sheet (Supplementary Fig. 9a) and shined UV light on the sample for 5 minutes before removing the mask (Supplementary Fig. 10). In the unmasked region, the UV light released the ATP from the blocking group and activated the fluid by allowing the ATP to fuel the local kinesin motors; in the masked region, the fluid remained quiescent (Fig. 1c; Supplementary Video 1).[45] To explore how the progression exponent changed with active fluid bulk mean speed, we accelerated fluid flows by making the flow cell taller by stacking layers of Parafilm to decrease hydrodynamic resistance (Fig. 4a inset right).[48,52] To explore how the spatial nonuniformity of activity influenced the mixing efficacy of the active-inactive fluid system, we masked the sample with checkerboard-patterned masks (FineLine Imaging, Fig. 7a).

**Image samples with dual fluorescent channels.** We imaged the active fluids with epifluorescent microscopy (Ti2-E Inverted Microscope, Nikon, MEA54000) with the commercial image acquisition software Nikon NIS Elements version 5.11.03. To capture a wide area of the sample ($20 \times 4$ mm$^2$), we used a 4× objective lens (CFI Plan Apo Lambda 4× Obj, Nikon, MRD00045, NA 0.2) to image 3 to 4 adjacent frames rapidly ($\lesssim$3 s) and stitched the micrographs into one large image for flow and dye dispersion analyses (Figs. 1, 2, 4, 5, & 7; Supplementary Fig. 9b).

Performing these analyses required monitoring at least two components in two fluorescent channels in each sample; for example, the dye dispersion experiments (Fig. 5) required analyzing fluorescent dyes (excitation: 490 nm; emission: 525nm) and Flash Red-labeled tracers (excitation: 660 nm; emission: 690 nm) simultaneously. This could have been accomplished by programming a microscope to rapidly switch back and forth between filter cubes, but this would have quickly worn down the turret motor and the time required to switch filter cubes ($\gtrsim$4 s) and move the stage to capture adjacent images and stitch them (~3 s) would have made the minimum time interval between frames $\gtrsim$10 s, which would have prevented us from tracking high-density tracers (1000 mm$^{-3}$ with a mean separation of 5 µm in a 0.1-mm-thick sample) whose speeds were 1 to 10 µm/s, even with a predictive Lagrangian tracking algorithm.[69] To overcome this technical challenge in imaging our samples, we established a dual-channel imaging system that consisted of a multiband pass filter cube (Multi LED set, Chroma, 89402–ET) and voltage trigger (Nikon) placed between the light source (pE-300$^{ultra}$, CoolLED, BU0080) and camera (Andor Zyla, Nikon, ZYLA5.5-USB3). Instead of changing filter cubes, the multiband pass filter cube allowed us to switch between multiple emission and excitation bands by switching between channels with the same filter cube (Supplementary Fig. 11). We alternatively activated the blue (401–500 nm) and red (500–700 nm) LEDs to excite and observe the fluorescent dyes and tracers almost simultaneously. The LED light source communicated directly with the camera via voltage triggering to coordinate LED activation time and bypass computer control to further boost the light switching rate. This technique shortened our channel switching time to 3 to 5 µs; thus the time interval between image acquisitions of different fluorescent channels was only limited by exposure times of each channel. This setup allowed us to image two fluorescent channels almost simultaneously (within milliseconds) and thus enabled us to monitor two fluorescent components side-by-side, such as microtubules and tracers (Fig. 1c; Supplementary Video 1), caged fluorescent dyes and tracers (Fig. 5c), and caged fluorescent dyes and microtubules (Supplementary Video 5).

**Analyze positions of active-inactive fluid interfaces.** We characterized the mixing kinematics of active and inactive fluids by analyzing the interface progression exponents $\gamma$ and coefficients $P_I$ (Figs. 2–4). These analyses required identification of the interface positions. We determined the interface positions by first tracking tracers in sequential images with the Lagrangian tracking algorithm,[69] which revealed the tracers' trajectories $r_i(t)$ and corresponding instantaneous velocities $v_i \equiv d\boldsymbol{r}_i/dt$. Then we analyzed the speed profile of tracers by binning the tracer speed $|v_i|$ across the width of the channel $S(x_j) \equiv \langle |v_i| \rangle_{i \in \mathrm{bin}\, j}$ where $x_j$ was the horizontal $x$ coordinate of the $j$th bin and the $\langle |v_i| \rangle_{i \in \mathrm{bin}\, j}$ represented the average speed of tracers in the $j$th bin. Then we normalized the speed profile by rescaling the speed profile to be 1 in the active zone and 0 in the inactive zone: $S^*(x) \equiv [S(x) - s_{in}]/[s_a - s_{in}]$, where $s_a$ is the average of speed profiles in the active zone and $s_{in}$ is the average of speed profiles in the inactive zone (Fig. 1e). Then we defined the width of the active-inactive fluid interface as where the normalized speed profile is between 0.2 and 0.8 (dashed lines in Fig. 1e inset) and the position of the interface $x_I$ as where the normalized speed profile is 0.5 (solid line). The interface position was then analyzed for each frame, which allowed us to determine the interface displacement $\Delta x \equiv x_I(\Delta t) - x_I(0)$ vs. time $t$ (Fig. 2a inset) and the squared interface displacement $\Delta x^2$ vs. time $t$ (Fig. 2a). To determine the interface progression exponent, $\gamma$, we fit $\log(\Delta x^2)$ vs. $\log(t)$ data to $\log(\Delta x^2) = \log(2P_I) + \gamma \log(t)$, with $P_I$ and $\gamma$ as fitting parameters (Figs. 2b & 4). To determine the interface progression coefficient, $P_I$, we assumed $\gamma = 1$ and fit $\Delta x^2$ vs. $t$ data to $\Delta x^2 = 2P_I t$ with $P_I$ as the fitting parameter (Figs. 2c, 3c, & 3d).

**Generate flow speed map of active-inactive fluid system.** To visualize the activity distribution in our active-inactive fluid system, we analyzed the distribution of flow speed to generate flow speed maps (Fig. 1d). To complete this analysis, we analyzed the flow velocities of fluids by analyzing the tracer motions in sequential images with a particle image velocimetry algorithm,[70] which revealed the flow velocity field $\boldsymbol{V}(\boldsymbol{r}, t)$ and associated distribution of flow speed $|\boldsymbol{V}(\boldsymbol{r}, t)|$ in each frame. A heat color bar (Fig. 1d) was used to plot the speed distributions into color maps to reveal the evolution of speed distribution from the pre-activated state (black) to the homogeneously activated state (red/yellow).

**Numerically solve the Fick's law equations.** We modeled diffusion-dominated active-inactive fluid mixing with the Fick's law equation, which required us to solve for the concentration of ATP (Eq. 1). To simplify the modeling, we considered a one-dimensional active fluid system confined in a segment $x = 0$–$L$ where $L = 20$ mm, the length of our experimental sample. Given that ATP is confined in the segment, we applied no-flux boundary conditions to the ATP concentrations (Eq. 2). In the experiment, we exposed the left side of the sample to UV light to release ATP, so in our model the ATP concentration has a step function as its initial condition (Eq. 3). With the initial condition and boundary conditions, we solved the Fick's law equation to determine the spatial and temporal distribution of ATP. We used Mathematica 13.0 to numerically solve this differential equation with the NDSolveValue function by feeding Eqs. 1–3 into the function followed by specifying the spatial and temporal domains, which output the numerical solution of ATP concentration $C(x, t)$ and allowed us to determine the evolution of ATP distribution (Fig. 3a). Then we converted the ATP distribution to mean speed distribution of active fluid by the Michaelis-Menten kinetics (Eq. 4; Fig. 3b inset), which showed a uniform mean speed in the left side of the sample followed by gradual activation of the right side of the system until a uniform state was reached (Fig. 3b). Then we defined the position of the active-inactive fluid interface $\Delta x$ as where the mean speed decayed to a half (Fig. 3c inset), which allowed us to plot the squared interface displacement as a function of time (Fig. 3c). The plot in log-log axes exhibited a line with a unit slope, which suggested that the squared interface displacement is linearly proportional to time. By assuming this linear relation, we determined the interface progression coefficient $P_I$ by fitting $\Delta x^2 = 2P_I t$ with $P_I$ as the fitting parameter. Finally, we repeated the calculation with different initial ATP concentrations in the left side of the system, $C_0$, and plotted the interface progression coefficient, $P_I$, as a function of $C_0$. This plot allowed us to compare the simulation results with the experimental measurements to examine the validity of our Fick's law-based model (Fig. 3d).

**Numerically solve the active nematohydrodynamic equations in weak forms.** To model the mixing of active and inactive fluids, we adopted Varghese *et al.*'s active fluid model to include the dynamics of the ATP concentration field.[22] Our model had four main components: (1) the kinetic equation describing the kinematics of self-elongating rods that flow and orient with the solvent as well as diffuse translationally and rotationally (Eq. 6), (2) the Stokes equation describing how the solvent was driven by the active stress exerted by the self-elongating rods (Eq. 7), (3) the relation between $\alpha$ and ATP that describes how the active stress depended on the nonuniform ATP distribution (Eq. 8), and (4) the continuity equation of ATP transport that describes how ATP diffuses as well as flows with the solvent (Eq. 9). We numerically solved these coupled equations with appropriate boundary and initial conditions using the finite element method by first converting them to their weak forms and then implementing them symbolically in COMSOL Multiphysics™.[60] We show below the weak form of the convection-diffusion equation governing the evolution of ATP concentration field:

$$\int_\Gamma dx^* \, dy^* \, \widetilde{T} \frac{\partial C}{\partial t^*} = - \int_\Gamma dx^* \, dy^* \, D^* \left( \frac{\partial \widetilde{T}}{\partial x^*} \frac{\partial C}{\partial x^*} + \frac{\partial \widetilde{T}}{\partial y^*} \frac{\partial C}{\partial y^*} \right) - \int_\Gamma dx^* \, dy^* \, \widetilde{T} \left( u_x^* \frac{\partial C}{\partial x^*} + u_y^* \frac{\partial C}{\partial y^*} \right),$$  10

where $\widetilde{T}(x^*, y^*)$ is the test function and $\Gamma$ represents the system spatial domain. After solving these equations, we determined the spatial and temporal distributions of ATP concentrations and active fluid flow speeds (Fig. 6a and Supplementary Video 4), which allowed us to explore how the activity level of active fluid and molecular diffusion of ATP influenced the mixing process of ATP in nonuniform active fluid systems (Figs. 6b & c).

**Simulate dispersion of checkerboard-patterned ATP in active and inactive fluids.** We applied our established hydrodynamic model to simulate how the initially checkerboard-patterned ATP would disperse in active fluid and inactive fluid (Fig. 8). In this exploration, we used the same equations for $\boldsymbol{Q}$, $\boldsymbol{u}$ and $C$ (Eqs. 6-9), along with their initial and boundary conditions, except that $C$ was initialized in a checkerboard pattern in a $45 \times 45$ simulation box. We used two different checkerboard patterns for different trials. One has the $xy$-axis origin in the center of a grid square:

$$C(x, y, t = 0) = C_0 \, \text{mod} \left\{ \text{mod} \left[ \text{ceil} \left( \frac{x^*}{a^*} - 0.5 \right), 2 \right] + \text{mod} \left[ \text{ceil} \left( \frac{y^*}{a^*} - 0.5 \right), 2 \right] + 1, 2 \right\},$$  11

where $C_0 = 5$ mM is the initial ATP concentration, $a^*$ represents the dimensionless grid size of the checkerboard pattern, $\text{mod}(i, j)$ represents the remainder of $i$ divided by $j$, and $\text{ceil}(x)$ represents the rounding of $x$ toward positive infinity (e.g., Fig. 8a with $a^* = 22$). The other checkerboard pattern had the $xy$-axis origin at a vertex of the grid:

$$C(x, y, t = 0) = C_0 \, \text{mod} \left\{ \text{mod} \left[ \text{ceil} \left( \frac{x^*}{a^*} \right), 2 \right] + \text{mod} \left[ \text{ceil} \left( \frac{y^*}{a^*} \right), 2 \right] + 1, 2 \right\}.$$  12

The simulation was performed for the dimensionless grid size $a^*$ ranging from 2 to 22 for both active ($a^* = 25$) and inactive ($a^* = 0$) systems (Fig. 8a). We analyzed the corresponding mixing time (averaged over 2 trials) as a function of $a^*$ (Fig. 8b) to reveal how the $a^*$-dependence of mixing times differed in active and inactive fluid systems.

**Data availability.** The data that support the findings of this study are available from the corresponding author upon reasonable request.

**Code availability.** The Mathematica script used to solve the Fick's law-based model (Eqs. 1-4) and the COMSOL file used to solve the active nematohydrodynamic equations coupled with active transport of ATP (Eqs. 6-9) has been deposited on figshare (https://doi.org/10.6084/m9.figshare.20332806.v1).

**Acknowledgements**

T.E.B. and K.-T.W. would like to thank Drs. Seth Fraden and Aparna Baskaran of Brandeis University for insightful discussions on experiments and modeling in this manuscript. T.E.B. and K.-T.W. would also like to thank Dr. John Berezney of Brandeis University for assisting us in developing the UV light setup (Supplementary Fig. 10). We thank Ellie Lin from Lin Life Science for her assistance on editing the manuscript to enhance its flow and readability. T.E.B., E.H.T., and K.-T.W. acknowledge support from the National Science Foundation (NSF-CBET-2045621). This research was performed with computational resources supported by the Academic & Research Computing Group at Worcester Polytechnic Institute. We acknowledge the Brandeis Materials Research Science and Engineering Center (NSF-MRSEC-DMR-2011846) for use of the Biological Materials Facility. Parts of the work by C.-C. C. were funded and supported through the National Science and Technology Council (NSTC), Taiwan, under grant No. 111-2221-E-006-102-MY3, and through the 2022 Early Career Award from the College of Engineering and the Headquarters of University Advancement at National Cheng Kung University, which was sponsored by the Ministry of Education, Taiwan. M.M.N. was supported by the U.S. Department of Energy, Office of Science, Office of Basic Energy Sciences under Award No. DE-SC0022280.

**Author Contributions**

T.E.B., M.E.V., and K.-T.W. performed the research and designed the experiments; M.E.V. initiated the experiments; T.E.B., M.E.V., E.H.T. and J.H.D. collected experimental data; T.E.B., M.E.V., and K.-T.W. organized and analyzed the data; T.E.B., E.H.T., C.-C.C., M.M.N. and K.-T.W. established the continuum simulation platform on modeling nonuniform active fluid systems; T.E.B. and K.-T.W. wrote the manuscript; and K.-T.W. supervised the research. All authors reviewed the manuscript.

**Additional Information**
**Competing interests statement.** The authors declare that they have no competing interests.

**Correspondence.** Correspondence and requests for materials should be addressed to K.-T.W. (kwu@wpi.edu). Active fluid simulation questions should be addressed to C.-C.C. (chuehcc@mail.ncku.edu.tw), M.M.N. (mmnsps@rit.edu), and K.-T.W. (kwu@wpi.edu).

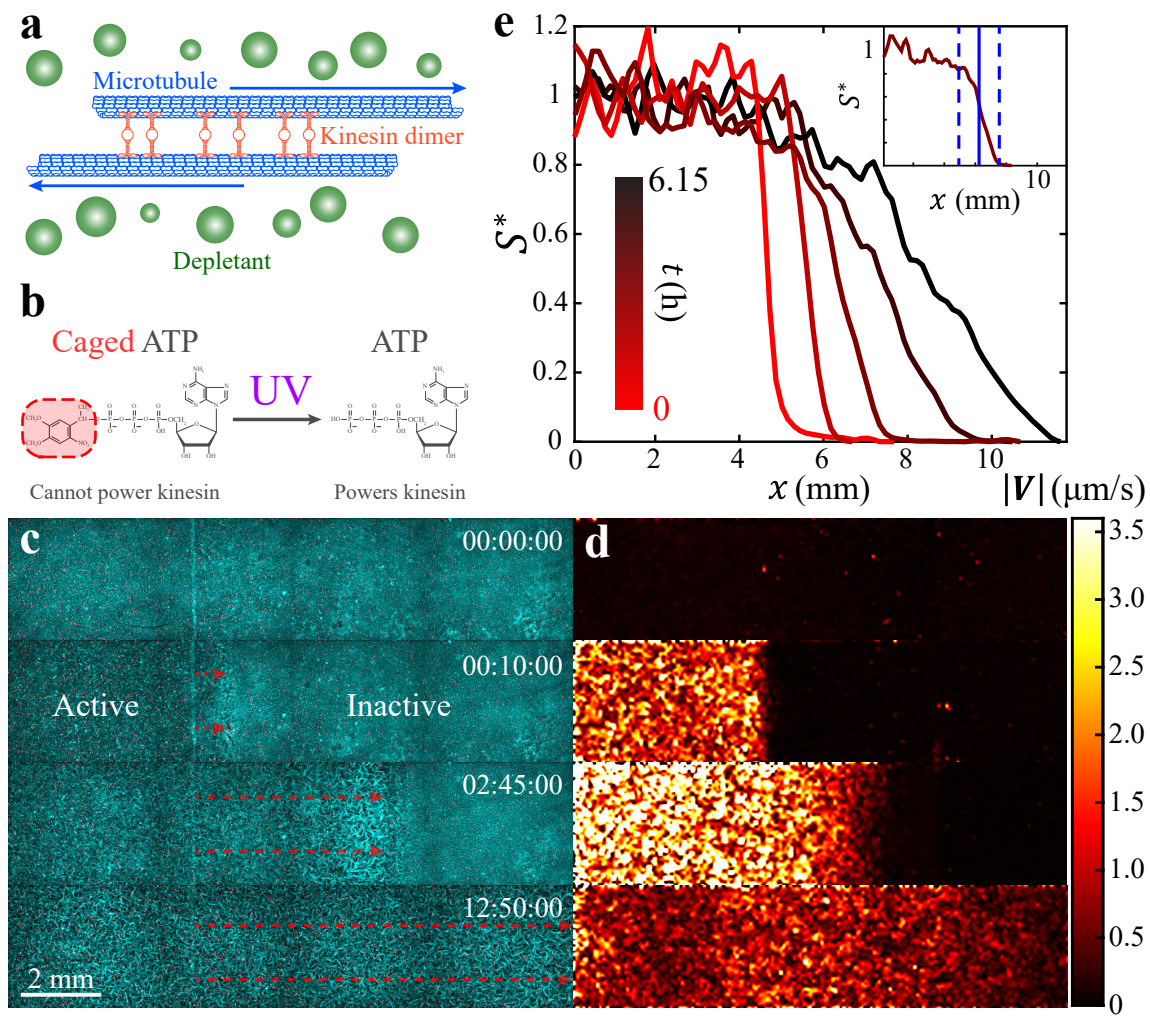

Fig. 1: (Experimental results) Mixing of activated and inactive microtubule-kinesin fluid. (a) Microscopic dynamics in microtubule-kinesin active fluid. Depletants force microtubules into bundles where the microtubules can be bridged by kinesin motor dimers. The kinesin motors "walk" along the microtubules, forcing them to slide apart. The collective sliding dynamics cause the microtubules to form an extensile microtubule network that stirs the surrounding solvents and causes millimeter-scale chaotic flows.[14] (b) To develop an experimental system where we could create a distinct boundary between active and inactive fluid, we synthesized microtubule-kinesin active fluid with caged ATP. The caged ATP is not hydrolysable by kinesin motors, and thus cannot power the active fluid, until it is released by exposure to ultraviolet light.[45,46] This process is not reversible. (c) We exposed only one side of the sample to ultraviolet light, which released the ATP and activated the microtubule-kinesin mixture on that side of the channel. The released ATP dispersed toward the unexposed region, which activated the inactive fluid and expanded the active region until the system reached an activity-homogeneous state (Supplementary Video 1). Because of the limited speed of multi-position imaging, only one-quarter of the active region was imaged. (d) Tracking tracer particles revealed the speed distribution of fluid flows, showing the activation of the left-hand side by UV light and the expansion of the active region into the inactive region. (e) Binning the same-time speeds vertically across the interface of active and inactive fluids revealed the speed profile $S$ which is normalized as $S^*(x) \equiv [S(x) - s_{in}]/[s_a - s_{in}]$, where $s_a$ is the average of speed profiles in the active zone and $s_{in}$ is the average of speed profiles in the inactive zone. Inset: The interface of the active and inactive fluids is determined as the region where the normalized speed profile is between 0.2 and 0.8 (dashed blue lines). The position of the interface is determined as where the normalized speed profile was 0.5 (solid blue line).

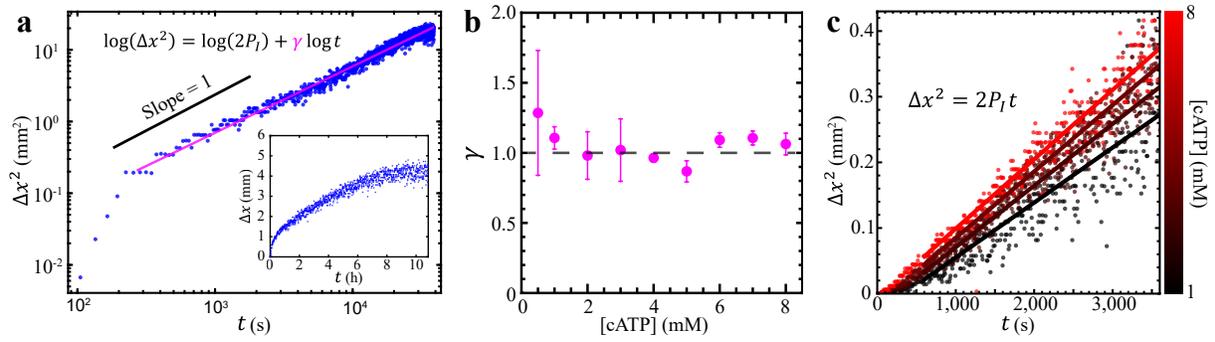

Fig. 2: (Experimental results) The progression of the active-inactive interface was governed by a diffusion-like process of ATP at the interface, regardless of caged ATP concentration. (a) Squared interface displacement ($\Delta x^2$) as a function of time ($t$) revealed a long-term ($t \gtrsim 200$ s) linear relation: $\Delta x^2 \sim t^{\gamma}$ with the interface progression exponent $\gamma \approx 1$. Inset: The interface displacement versus time showed that the interface moved rapidly initially and then gradually slowed down. (b) The interface progression exponent was $\gamma \approx 1$ on average and was independent of the caged ATP concentration. Each error bar represents the standard deviation of $\geq 3$ trials. (c) Selected examples of squared interface displacement versus time for four caged ATP concentrations from 1 mM (black) to 8 mM (red). The progression rate of the interface was characterized with an interface progression coefficient $P_I$ determined by fitting the $\Delta x^2$ vs. $t$ data to $\Delta x^2 = 2P_I t$ with $P_I$ as the fitting parameter (colored lines). Increasing caged ATP concentrations increased the interface progression coefficient (steeper fit lines from black to red), which indicates that the interface progressed more quickly at higher caged ATP concentrations.

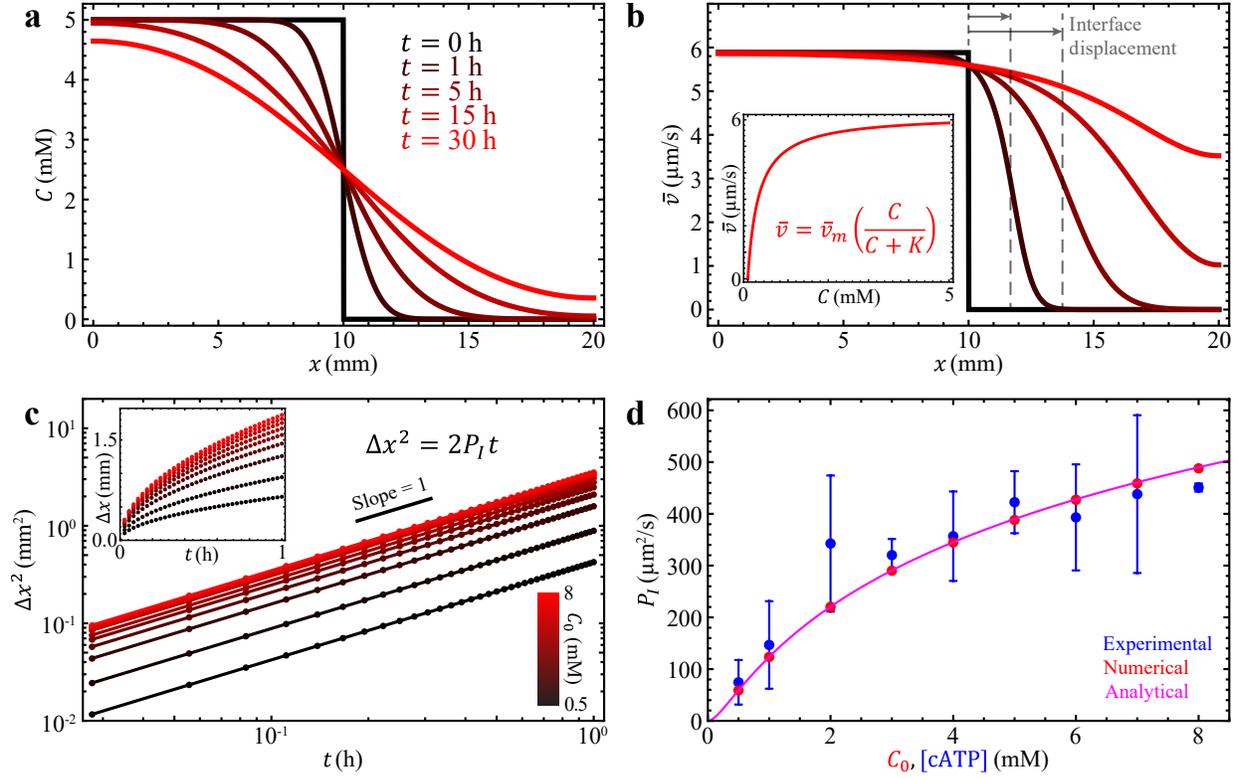

Fig. 3: (Modeling results) Fick's law of diffusion and Michaelis-Menten kinetics captured the diffusion-like mixing of active and inactive fluids. (a) The simulated distribution of ATP concentrations started as a step function (black, $t = 0$ h) and then developed into a smoothed hill function (red, $t = 30$ h) as ATP evolved from a one-sided distributed to a homogeneous state. (b) The model converted the ATP distribution into the speed distribution of active fluid via Michaelis-Menten kinetics: $\bar{v} = \bar{v}_m[C/(C + K)]$, where $\bar{v}_m = 6.2$ μm/s and $K = 270$ μM (based on our previous studies[48]). The corresponding mean speed distribution of active fluid evolved from a step function distribution (black, $t = 0$ h) to a near-constant function (red, $t = 30$ h) (Supplementary Video 2). Inset: The plot of the Michaelis-Menten equation (Eq. 4). (c) In the simulation, the diffusion-driven mixing process led the squared interface displacement to be proportional to time, regardless of initial ATP concentration $C_0$ (see Supplementary Discussion 3 for derivation of $\Delta x^2 \propto t$). Inset: Interface displacement increased rapidly with time initially, followed by a gradual deceleration similar to the experimental observation (Fig. 2a inset). (d) In the simulation, the interface progression coefficient $P_I$ was determined by fitting the $\Delta x^2$ vs. $t$ data (Panel c) to $\Delta x^2 = 2P_I t$ with $P_I$ as fitting parameter. The model $P_I$ increased with the initial concentration of ATP, $C_0$ (red dots), similarly to how the experimentally analyzed $P_I$ varied with caged ATP concentration (blue dots; each error bar represents the standard deviation of $\geq 3$ trials). The model $P_I$ and experimental $P_I$ differed by only ~10%. The magenta curve shows the analytical solution, $P_I(C_0)$ (Eq. S7), which reproduced the numerical results (red dots). (See Supplementary Discussion 3 for derivation of $P_I$ as a function of $C_0$.)

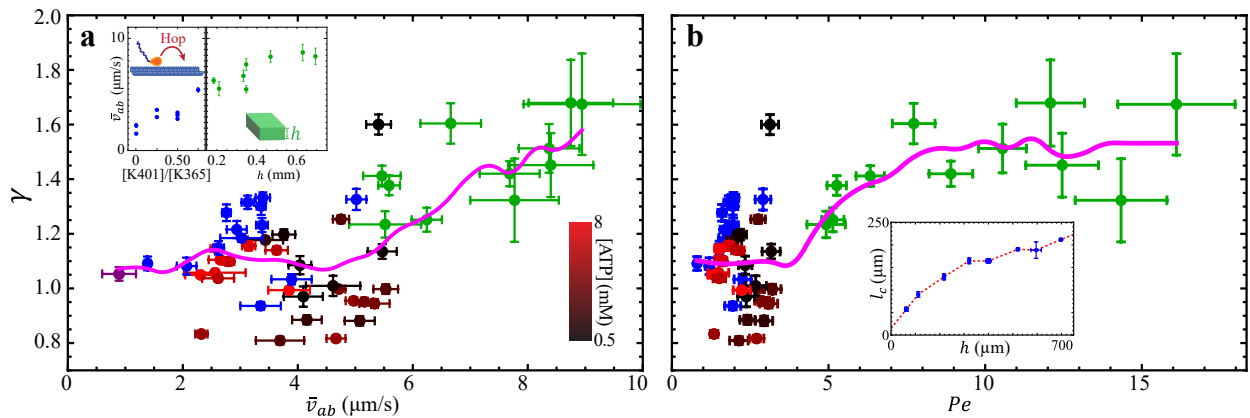

Fig. 4: (Experimental results) Progression of the active-inactive interface transitioned from diffusion-like ($\gamma \approx 1$) to superdiffusion-like ($\gamma > 1$) as the active transport changed from diffusion-dominated ($Pe \lesssim 3$) to convection-dominated ($Pe \gtrsim 3$). (a) The active-inactive interface progression exponent ($\gamma$) increased with the flow-speed level of the active fluid ($\bar{v}_{ab}$). Shown are data from experiments with low ATP concentration (0.5 mM, black dots), high ATP concentration (8 mM, red dots), decreased flow speeds (from nonprocessive motors partially replacing processive motors; blue dots), increased flow speeds (from increased sample height; green dots),[48] and both nonprocessive motors and increased sample height (purple dot). The magenta curve represents the moving average of $\gamma$. Although the analyzed $\gamma$ from each experiment was noisy, the moving averaged $\gamma$ exhibited an overall monotonic increase with the flow-speed level of active fluid $\bar{v}_{ab}$. Each dot represents one experimental measurement. Each error bar in $\gamma$ represents the slope fitting error in $\ln \Delta x^2 = \ln(2P_l) + \gamma \ln t$ (Fig. 2a), and each error bar in $\bar{v}_{ab}$ represents the standard deviation of flow speeds in the active region. Inset: The flow-speed level of the active fluid was tuned by replacing processive motors (K401) with nonprocessive motors (K365) with the same overall motor concentrations (120 nM) (left)[36,48] or by altering the sample height (right).[48,52] (b) The same data as in Panel a, plotted as a function of Péclet number, $Pe \equiv \bar{v}_{ab} l_c / D$, where $l_c$ is the correlation length of flow velocity in active fluid deduced from sample container height $h$ (inset) and $D$ is the diffusion coefficient of ATP. Each error bar in $\gamma$ is the same as in Panel a, and each error bar in $Pe$ represents propagated uncertainties from $\bar{v}_{ab}$ in Panel a and $l_c$ in inset. Inset: Correlation length of flow velocity in active fluid $l_c$ increased monotonically with sample container height $h$.[39] The red dashed line represents the line interpolation of blue dots. The error bars represent the standard deviations of two trials. (See Supplementary Discussion 4 for measurements and analyses of $l_c$.)

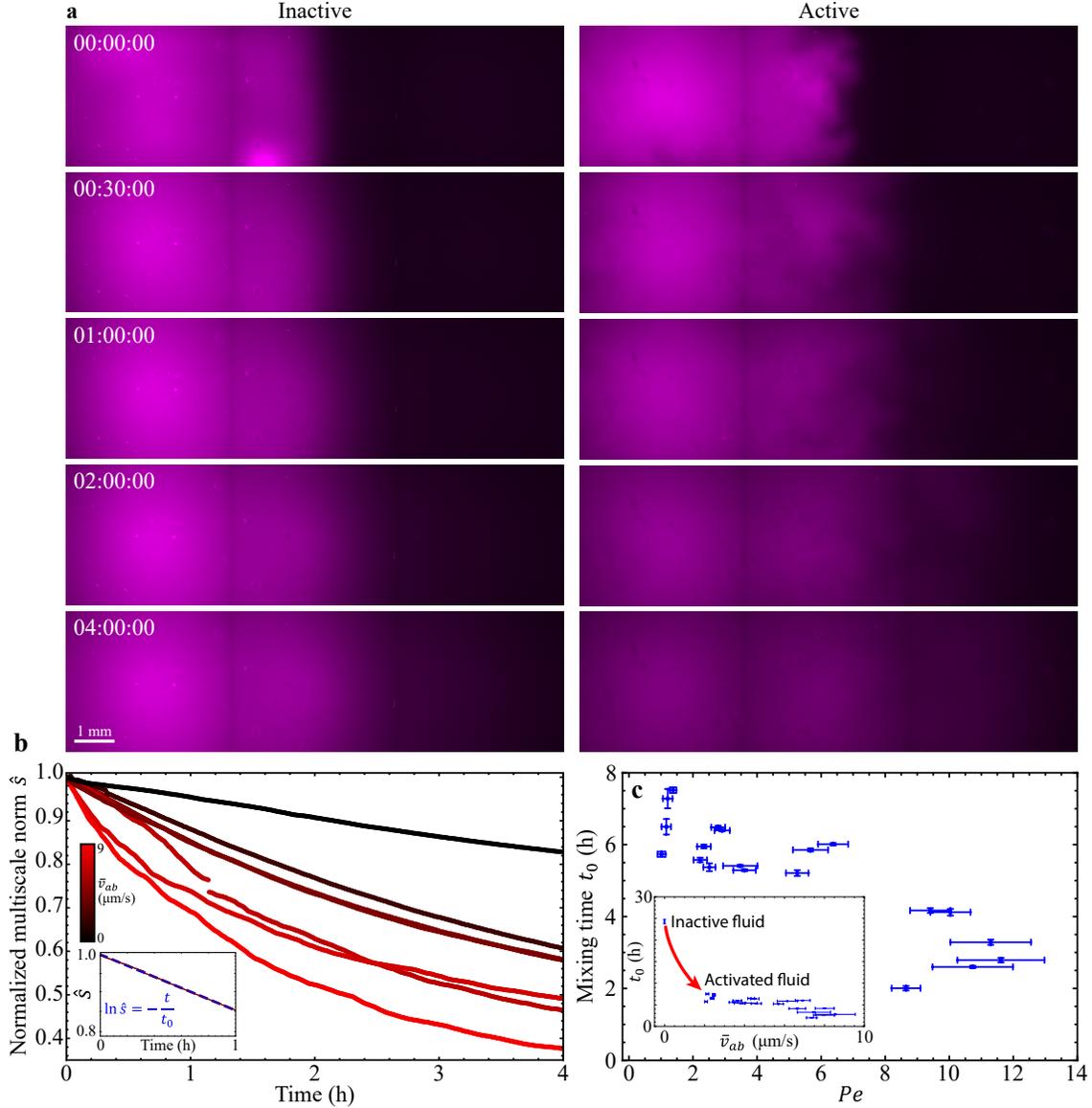

Fig. 5: (Experimental results) Active fluid flows promoted mixing of UV-activated fluorescent dyes, which were initially activated in the left-hand side of the container only. (a) Dispersion of UV-activated fluorescent dyes (magenta) in inactive ($\bar{v}_{ab} = 0$; left column) and active ($\bar{v}_{ab} = 8.2$ μm/s; right column) microtubule-kinesin fluid. Active fluid flows actively transported fluorescent dyes and enhanced their dispersion. Time stamps are hour:minute:second. (See also Supplementary Video 3.) (b) Selected examples of normalized multiscale norm vs. time for different active bulk flow speeds, $\bar{v}_{ab}$. Normalized multiscale norm $\hat{s}(t)$ decreased faster in a faster-flowing active fluid system. Inset: The normalized multiscale norm, $\hat{s}(t)$, in log-linear axes behaved as a straight line, which suggests that the norm decayed exponentially with time. The decay time scale $t_0$ (or mixing time) was determined by fitting the normalized multiscale norm versus time data to $\ln \hat{s} = -t/t_0$ with $t_0$ as the fitting parameter (dashed blue line). (c) The mixing time decreased monotonically with Péclet number, which demonstrated that a stronger convection mechanism led to faster mixing of suspended components. Each dot represents one experimental measurement. Each error bar in $t_0$ represents the slope fitting error in $\ln \hat{s} = -t/t_0$ (Panel b inset), and each error bar in $Pe$ (defined as $Pe \equiv \bar{v}_{ab}l_c/D$) represents propagated uncertainties from $\bar{v}_{ab}$ (see inset) and $l_c$ (see Supplementary Fig. 5d). Inset: Mixing time, $t_0$, as a function of active bulk mean speed, $\bar{v}_{ab}$. Each error bar in $t_0$ is the same as in Panel c, and each error bar in $\bar{v}_{ab}$ represents the standard deviation of flow speeds in the active region. Notably, the mixing time of the inactive fluid system ($\bar{v}_{ab} = 0$) was 24 hours (top-left dot); minimally activating the fluid ($\bar{v}_{ab} = 2$ μm/s) reduced the mixing time to 8 hours.

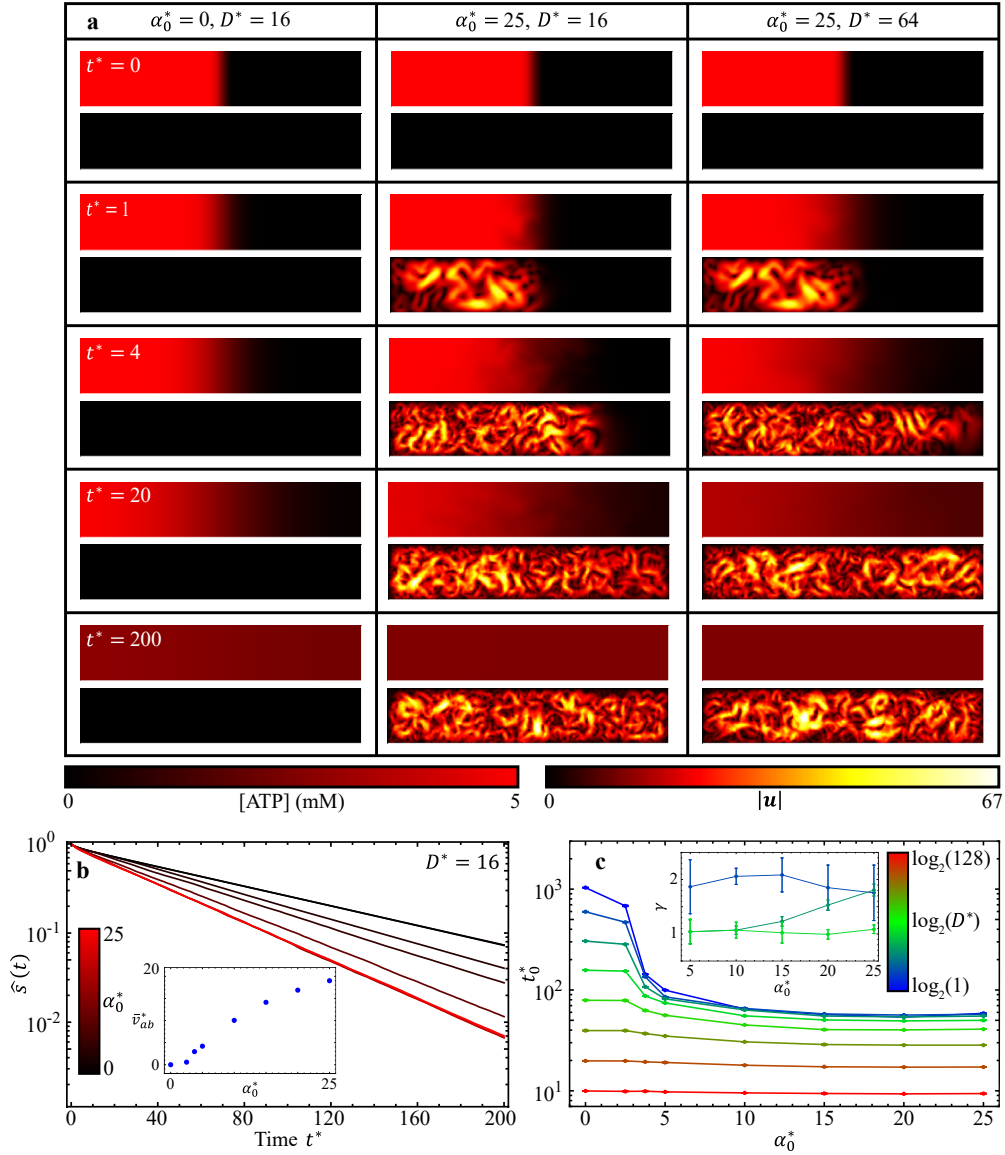

Fig. 6: (Modeling results) A continuous active fluid simulation revealed that the mixing time of ATP depended on the dimensionless molecular diffusion coefficient of ATP and the dimensionless activity level of active fluid. (a) Table of ATP concentration (top panels) and active fluid flow speed (bottom panels) maps for various dimensionless activity levels $\alpha_0^*$ and molecular diffusion coefficients $D^*$. When the fluid had no activity ($\alpha_0^* = 0$; left column), ATP dispersed to the right side of the system only by molecular diffusion; the dispersion was enhanced when the active fluid started to flow and actively transport ATP ($\alpha_0^* = 25$; middle column). The dispersion was further enhanced when ATP diffused significantly faster ($D^* = 64$; right column) (Supplementary Video 4). (b) Evolution of normalized multiscale norm for $\alpha_0^* = 0$–25 while keeping $D^* = 16$. The normalized multiscale norms decayed exponentially with time: $\hat{s} = \exp(-t^*/t_0^*)$, where $t_0^*$ is the dimensionless mixing time. Inset: Dimensionless mean speed of active fluid in active region $\bar{v}_{ab}^*$ monotonically increased with dimensionless activity level $\alpha_0^*$. (c) Dimensionless ATP mixing times, $t_0^*$, as a function of dimensionless activity level, $\alpha_0^*$, for various dimensionless molecular diffusion coefficients, $D^*$. Increasing both $\alpha_0^*$ and $D^*$ decreased mixing time monotonically. Each error bar in $t_0^*$ represents the fitting error of $\hat{s}$ vs. $t^*$ to $\ln \hat{s} = -t^*/t_0^*$ (Panel b). Inset: Active-inactive interface progression exponent $\gamma$ as a function of dimensionless activity level $\alpha_0^*$ for dimensionless molecular diffusion coefficients $D^* = 2$ (dark blue), 4 (dark green), and 8 (light green). Each error bar in $\gamma$ represents the slope fitting error as in Fig. 2a. Increasing $D^*$ decreased $\gamma$ (from dark blue to light green curve), whereas increasing $\alpha_0^*$ increased $\gamma$ (dark green curve).

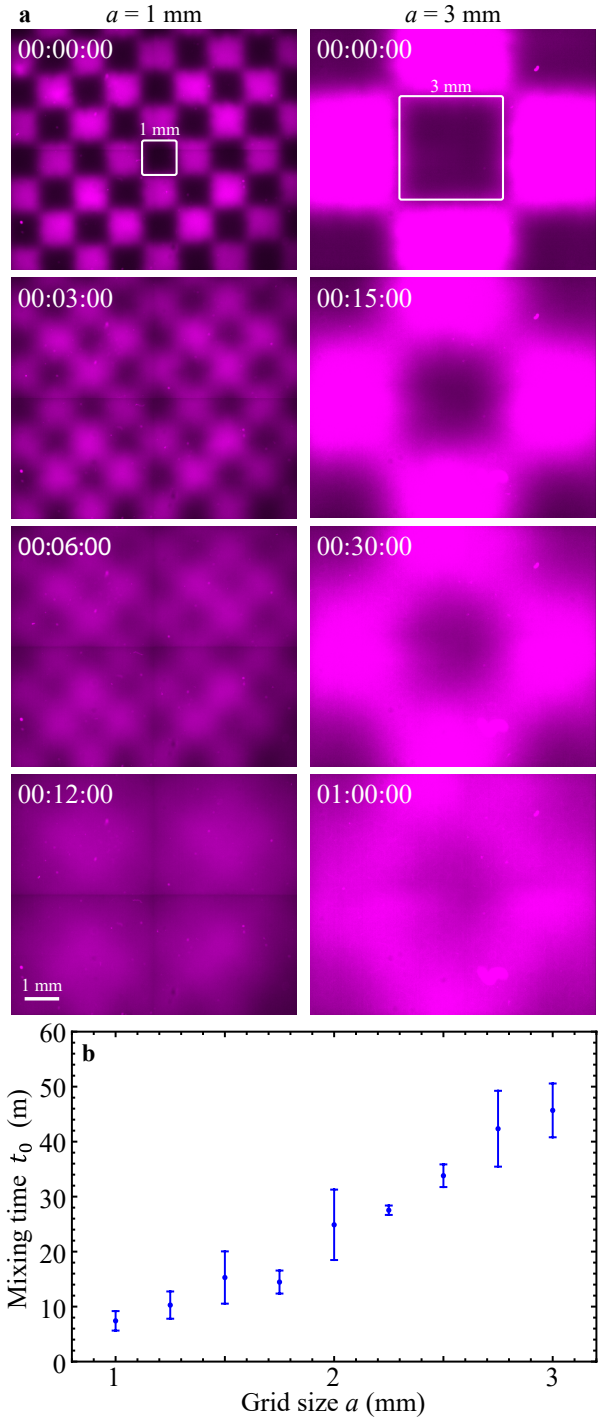

Fig. 7: (Experimental results) Fluid activated in a checkerboard pattern mixed faster when the checkerboard grid was smaller. (a) Checkerboard-patterned UV lights were used to activate active fluid and caged fluorescent dyes. (Supplementary Video 5). (b) The mixing time of the checkerboard-activated fluid increased with grid size, which demonstrated that the mixing efficacy of active fluid depended on distribution of activity: more nonuniform active fluid mixed the system more slowly. Each error bar in $t_0$ represents the standard deviation of two trials.

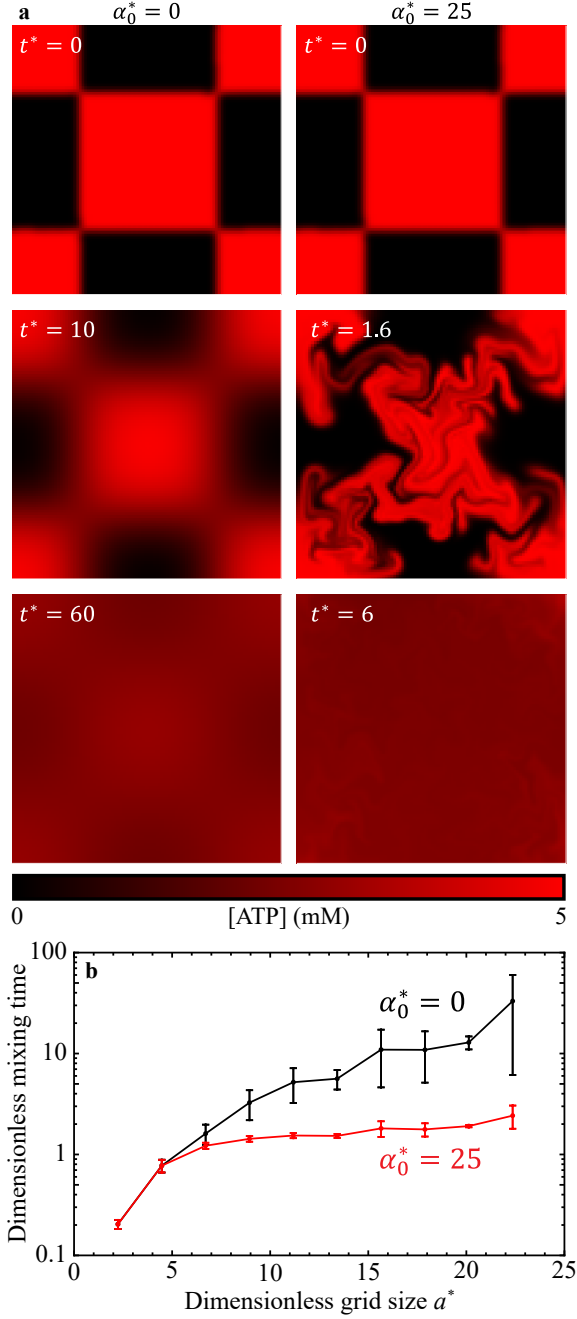

Fig. 8: (Modeling results) Simulations with ATP initially distributed in a checkerboard pattern in a $45 \times 45$ simulation box (Supplementary Video 6). (a) Distributions of ATP (red) dispersed from the checkerboard pattern (Eq. 11 with $a^* = 22$) to a homogeneous state by molecular diffusion only (left; $D^* = 1$, $\alpha_0^* = 0$) and by the combination of molecular diffusion and active fluid-induced convection (right; $D^* = 1$, $\alpha_0^* = 25$). (b) The dimensionless mixing time of ATP increased monotonically with the dimensionless checkerboard grid size $a^*$ in both active (red) and inactive (black) fluid systems. Error bars represent the standard deviation of two trials performed with different types of checkerboard pattern (Eqs. 11 & 12).

# Supplementary Information:

# Self-mixing in microtubule-kinesin active fluid from nonuniform to uniform distribution of activity

Teagan E Bate,[1] Megan E Varney,[2] Ezra H Taylor,[1] Joshua H Dickie,[1] Chih-Che Chueh,[3] Michael M Norton,[4] and Kun-Ta Wu[1,5,6,*]

[1]Department of Physics, Worcester Polytechnic Institute, Worcester, Massachusetts 01609, USA
[2]Department of Physics, New York University, New York, New York 10003, USA
[3]Department of Aeronautics and Astronautics, National Cheng Kung University, Tainan 701, Taiwan
[4]School of Physics and Astronomy, Rochester Institute of Technology, Rochester, New York 14623, USA
[5]Department of Mechanical Engineering, Worcester Polytechnic Institute, Worcester, Massachusetts 01609, USA
[6]The Martin Fisher School of Physics, Brandeis University, Waltham, Massachusetts 02454, USA
*Corresponding: kwu@wpi.edu

## Table of Contents

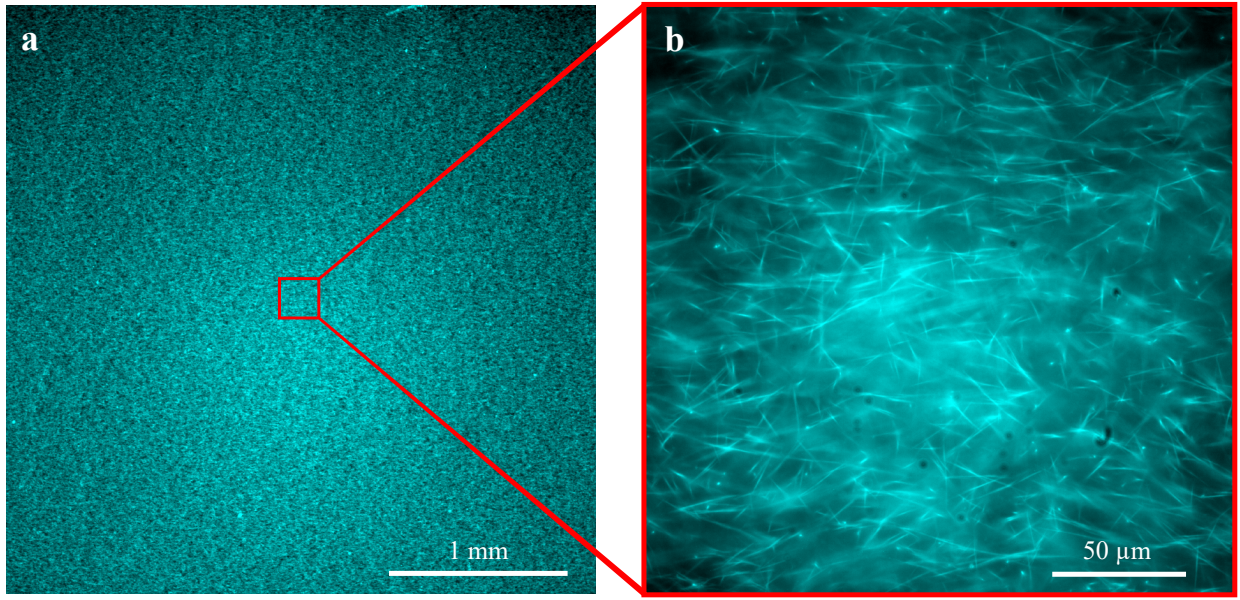

Supplementary Fig. 1: (Experimental results) Microtubules in inactive fluid loaded into the flow cell before UV activation. (a) Micrograph of microtubules at millimeter scale. (b) Micrograph of microtubules at micrometer scale, showing an initial dominant alignment of microtubules parallel to the long edge of the flow cell (horizontal direction) caused by the shear flow induced by pipetting the sample into the cell.[1]

**Supplementary Discussion 1: Estimation of ATP molecular diffusion coefficient in crosslinked microtubule networks**

ATP molecular diffusion plays an important role in our characterization of active-inactive fluid mixing. For example, we characterized the dominant mechanism in active transport (i.e., diffusion or convection) by introducing the Péclet number, $Pe$ (Figs. 4 & 5),[2,3] which depends on the molecular diffusion coefficient of ATP. We also adopted Fick's law (Eq. 1) to model the dispersion of ATP in low $Pe$ regimes (Fig. 3), which also requires the ATP diffusion coefficient. As such, it was important to determine the molecular diffusion coefficient of ATP. It is challenging to directly observe the diffusion of ATP, but fluorescein can be directly visualized, and thus we analyzed the molecular diffusion coefficient of fluorescein in the inactive microtubule network as the basis for estimating the molecular diffusion coefficient of ATP in our active fluid system.

We observed and characterized the dispersion of fluorescein near the interface of caged and UV-uncaged fluorescein suspended in inactive microtubule-kinesin fluid (Fig. 5a, Supplementary Fig. 2a) and compared the observation with the solution of Fick's law (Eq. 1) to extract the corresponding molecular diffusion coefficient. For each micrograph, we averaged the gray values vertically to determine the gray value profile (Supplementary Fig. 2b). We assumed that the fluorescein concentration was proportional to the gray values, and thus the profiles of gray values should satisfy the Fick's law equation (Eq. 1), whose solution in a boundless 1D system takes the form of a complementary error function:

$$C(x,t) = \frac{C_0}{2} \operatorname{erfc}\left(\frac{x - x_0}{2\sqrt{D(t - t_0)}}\right),$$

S1

where $C_0$ is the initial concentration of the suspended molecules, erfc is the complementary error function, and $D$ is the molecular diffusion coefficient of the suspended molecules. Thus, the profiles of gray values (G.V.) should take the form of

$$\text{G.V.}(x,t) = g_0 \operatorname{erfc}\left(\frac{x - x_0}{2\sqrt{\Gamma(t)}}\right) + g_0',$$

S2

where $g_0$ is the prefactor and $\Gamma \equiv D(t - t_0)$. We added the constant $g_0'$ to represent the gray value contributed from the background light in the microscope room. Then we fit this form to the profiles of gray values with $g_0$, $g_0'$, $x_0$, and $\Gamma$ as fitting parameters (Supplementary Fig. 2b inset). The fit $\Gamma$ is expected to be linear to time $t$, so we plotted $\Gamma$ as a function of time $t$ and fit $\Gamma$ vs. $t$ to $\Gamma = D(t - t_0)$ with $D$ and $t_0$ as fitting parameters (Supplementary Fig. 2c), which revealed that the molecular diffusion coefficient of fluorescein in our inactive microtubule-kinesin system was $D = 97.4 \pm 0.2$ μm²/s. This analyzed diffusion coefficient was one-fifth the value reported in aqueous solution (400-600 μm²/s).[4-7] We attribute this discrepancy to the rheologically complex environment of the inactive fluid that results from crosslinked microtubules.[8] We assume that the molecular diffusion of ATP molecules is similarly diminished in our experiments. The reported diffusion coefficient of ATP in water is 710 μm²/s,[9] and we therefore estimated the diffusivity of ATP to be 140 μm²/s in our models and calculations (Figs. 3-5).

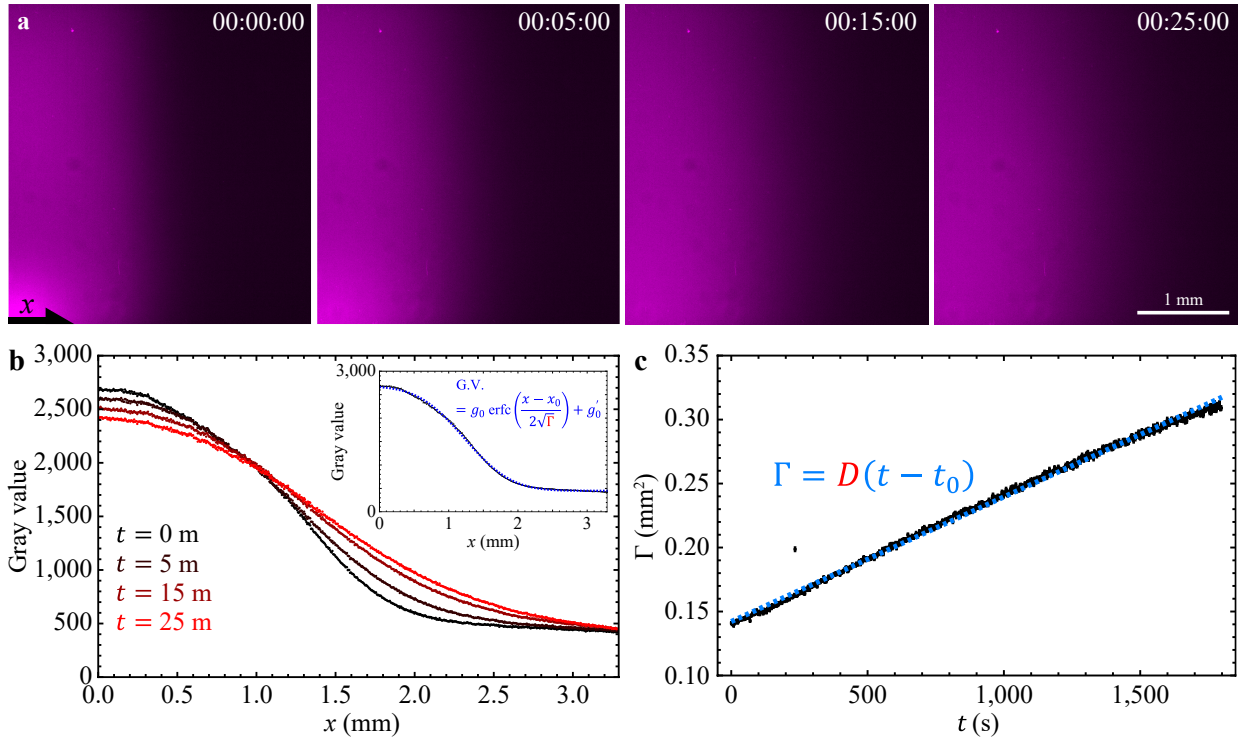

Supplementary Fig. 2: (Experimental results) Analysis of molecular diffusion coefficient of fluorescein suspended in inactive microtubule-kinesin fluid. The fluorescein was caged such that it only fluoresced after exposure to ultraviolet light (Fig. 5a). (a) Micrographs of fluorescein near the edge of the ultraviolet light-exposed area. The time stamps indicate hour:minute:second. (b) Profiles of gray values in the micrographs shown in Panel a. Inset: The gray value profile of the micrograph at time 00:00:00 (black dots) was fit to a complementary error function (Eq. S2; blue dashed curve) to extract the parameter $\Gamma$ which was expected to linearly increase with time $t$. (c) The fit parameter $\Gamma$ as a function of time $t$ (black dots) was fit to a line function (blue dashed line). The fit slope, $D = 97.4 \pm 0.2 \ \mu m^2/s$, represents the molecular diffusion coefficient of fluorescein suspended in crosslinked microtubule networks.

**Supplementary Discussion 2: Effect of flow speed-ATP relation on active-inactive interface progression in the Fick's law-based model**

We developed the Fick's law-based model (Fig. 3) to describe the experimentally observed mixing of active and inactive fluids (Fig. 2). Our model used Michaelis-Menten kinetics to convert distribution of ATP to flow speed (Eq. 4) because flow speed is driven by microtubule motion, which in turn depends on the stepping rate of kinesin motors which follows Michaelis-Menten kinetics.[10-12] Our previous work showed that Michaelis-Menten kinetics reasonably connect flow speed of active fluid with ATP concentrations when the ATP concentration is above 100 μM.[13] Below this concentration, inactive kinesin motor dimers start to act as a crosslinkers in the microtubule network, causing the network to behave more like an elastic gel, and Michaelis-Menten kinetics fail to describe the flow speed because Michaelis-Menten kinetics is an enzyme-based model that does not consider network rheology.

As such, our adoption of Michaelis-Menten kinetics to convert ATP concentration to flow speed is an approximation for active fluid with high ATP concentrations ($\gtrsim$100 μM), and it is unclear how the results of our Fick's law-based model would change if we chose a different relation between ATP concentration and flow speed. Here we considered that the flow speed is connected to ATP concentration via a positive power-law exponent, $n > 0$, of the Michaelis-Menten relation:

$$\bar{v} = \bar{v}_m \left( \frac{C}{C + K} \right)^n,$$ 
S3

where $\bar{v}_m = 6.2$ μm/s is the saturated mean speed of active fluid and $K = 270$ μM is the ATP concentration that leads to half of the saturated mean speed, $\bar{v}_m/2$ (Supplementary Fig. 3a).[13] Then we explored how $n$ would affect the predicted active-inactive interface progression in terms of progression exponent $\gamma$ and coefficient $P_I$.

To find $\gamma$ vs. $n$ and $P_I$ vs. $n$, we first considered the case of initial ATP concentration $C_0 = 8$ mM and solved the 1D Fick's law equation with the same boundary and initial conditions (Eqs. 1-3) for $C(x,t)$. Then we converted ATP concentration $C(x,t)$ to flow speed $\bar{v}(x,t)$ using our power-lawed Michaelis-Menten equation (Eq. S3), which showed that increasing power law exponents $n$ led to a slower progression of the active-inactive interface (Supplementary Fig. 3b). However, such an $n$-induced variation in interface progression appeared not to interfere with the relation of interface displacement with time; we found that the squared interface displacement increased linearly with time across our explored power-law exponents ($n = 0.5$-10; Supplementary Fig. 3c), which suggests that interface progression exponent $\gamma$ will equal 1 regardless of the value of the power-law exponent $n$ in our explored range of exponents (Supplementary Fig. 3d inset). Contrarily, we found that the interface progression coefficient decreased with increasing power-law exponents (Supplementary Fig. 3d). Overall, our exploration revealed that the active-inactive interface progression exponent being 1 was a consequence of the diffusion-like process of ATP dispersion. This result was insensitive to the choice of flow speed-ATP model (Eq. S3), but the interface progression coefficient varied rapidly with the model choice. Increasing the power-law exponent from 0.5 to 10 decreased the interface progression coefficient from 740 to 34 μm²/s. Given that our experimentally measured interface progression coefficient $P_I = 451 \pm 8$ μm²/s for 8 mM caged ATP concentration (Fig. 3d), this also suggested that selecting $n = 1$ (or slightly larger than 1) would best match our model with the experimental results (Supplementary Fig. 3d) and that Michaelis-Menten kinetics (Eq. 4) is a good approximation to connect ATP concentrations with the local flow speed of active fluid.

The calculations in our Fick's law-based model revealed that the interface progression coefficient for a given $C_0$ depends on the selected flow speed-ATP relation (Supplementary Fig. 3a), which implies that the ATP dependence of the interface progression coefficient $P_I(C_0)$ should change with the selected flow speed-ATP relation as well. To investigate, we repeated the above calculation and determined the interface progression coefficients, $P_I$, as a function of initial ATP concentration, $C_0$, for power-law exponents $n$ ranging from 0.5 to 10 (dots in Supplementary Fig. 4a). Our analysis revealed that increasing the power-law exponents decreased $P_I$ as a function of $C_0$ (Supplementary Fig. 4a) in a similar way that it did for flow speed (Supplementary Fig. 3a). Inspired by this observation, we fit each $P_I$ vs. $C_0$ to their corresponding power-lawed Michaelis-Menten equation with the same power-law exponent $n$:

$$P_I = P_m \left( \frac{C_0}{C_0 + K_P} \right)^n, \hspace{2cm} \text{S4}$$

with $P_m$ and $K_P$ as fitting parameters (curves in Supplementary Fig. 4a). The resulting data fit well to the equations, with overall goodness of fit $R^2 \geq 0.99$ (Supplementary Fig. 4b). This suggests that the ATP dependence in the flow speed could pass to the resulting interface progression coefficient $P_I$. This analysis also showed that the consistency between the model $P_I(C_0)$ and experimentally measured $P_I$ vs. $C_0$ (Fig. 3d) was under the condition of $n \approx 1$, which reenforced our assertion that adopting the Michaelis-Menten equation to convert ATP concentration to active fluid flow speed was an appropriate approach for building a coarse-grained model that matches the experimental results.

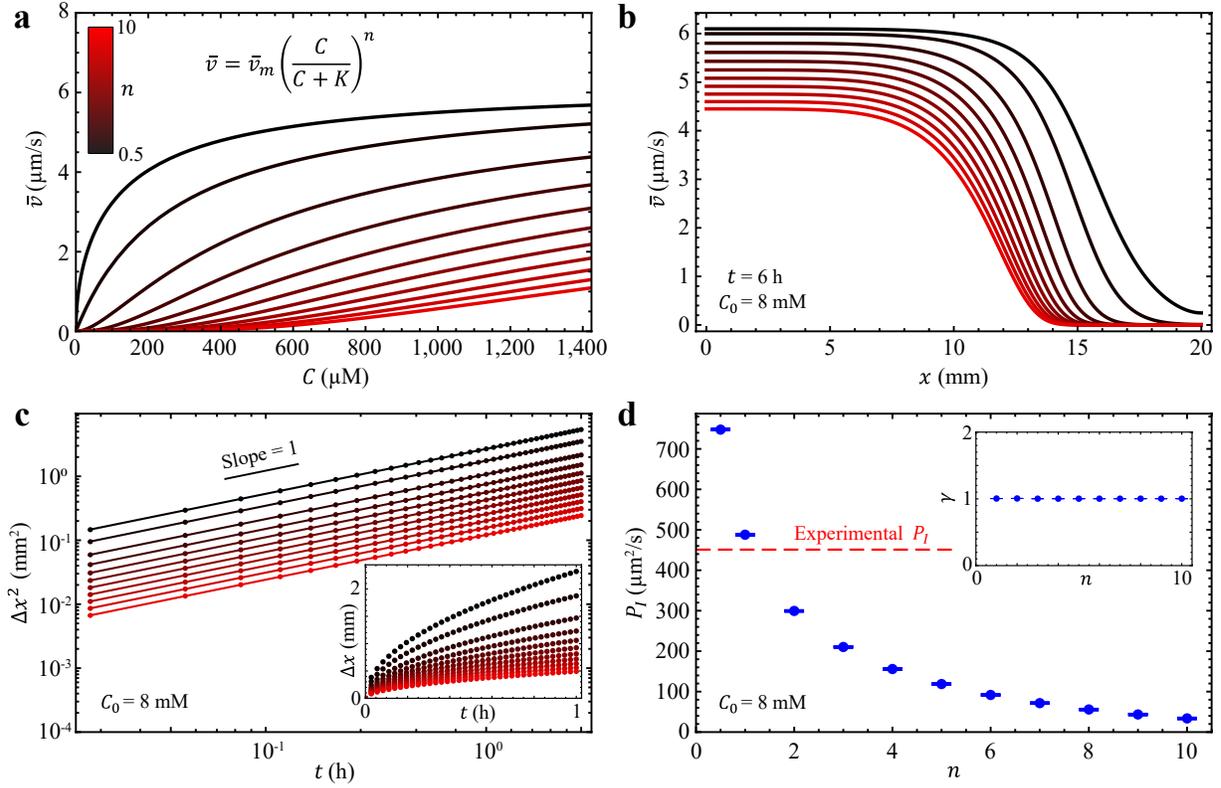

Supplementary Fig. 3: (Modeling results) The Fick's law-based model showed that the power-law exponent, $n$, in flow speed-ATP relation $\vec{v} = \vec{v}_m \{[\text{ATP}]/([\text{ATP}] + K)\}^n$, influenced the interface progression coefficients but not the interface progression exponents. (a) Plot of flow speed of active fluid as a function of ATP concentration, $\vec{v} = \vec{v}_m \{[\text{ATP}]/([\text{ATP}] + K)\}^n$, with different power-law exponents $n$, where $\vec{v}_m = 6.2$ μm/s and $K = 270$ μM.[13] (b) The distribution of flow speed at time $t = 6$ hours in the active fluid system for different power-law exponents. Each system had an initial ATP concentration of $C_0 = 8$ mM. The curves are colored based on the $n$ color bar in Panel a. (c) The corresponding squared interface displacement, $\Delta x^2$, increased linearly with time, $t$, for each explored power-law exponent $n$. The curves were the fitting of $\log(\Delta x^2) = \log(2P_I) + \gamma \log t$ with the interface progression coefficient, $P_I$, and interface progression exponent, $\gamma$, as fitting parameters. Inset: The corresponding interface displacement as a function of time for different power-law exponents. The dots and curves are colored based on the $n$ color bar in Panel a. (d) The corresponding interface progression coefficient, $P_I$, decreased with power-law exponent $n$. The red dashed line represents the experimentally measured $P_I = 451 \pm 8$ μm²/s for 8 mM caged ATP concentration (Fig. 3d). The error bars represent the fitting error in Panel c. Inset: The interface progression exponents remained 1 across explored power-law exponents $n = 0.5$–10. The error bars represent the fitting error in Panel c.

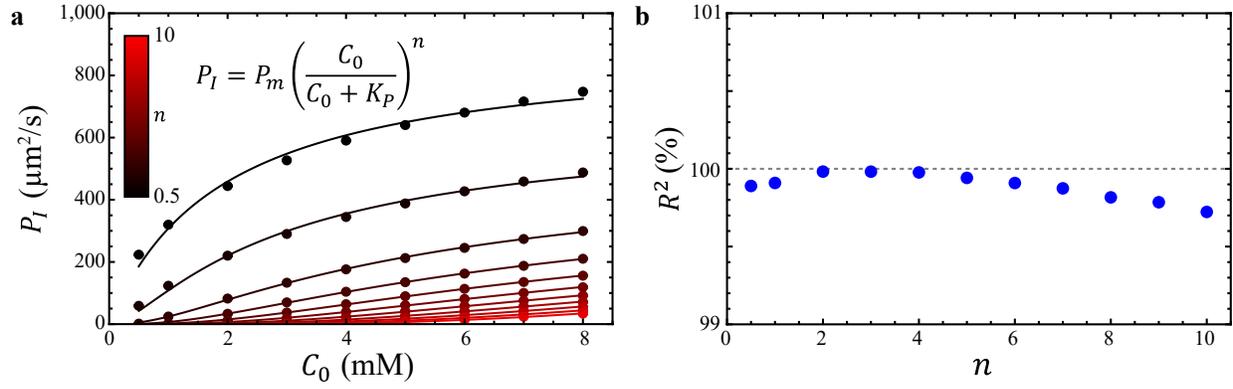

Supplementary Fig. 4: (Modeling results) The ATP dependence of interface progression coefficient $P_I(C_0)$ is inherent from the ATP dependence of the corresponding flow speed of active fluid $\bar{v}(C)$. (a) Interface progression coefficient, $P_I$, as a function of initial ATP concentration, $C_0$, for different power-law exponents $n$. The curves are the fitting to $P_I = P_m \{[\text{ATP}]/([\text{ATP}] + K_P) \}^n$ with $P_m$ and $K_P$ as fitting parameters. (b) The goodness of fit, $R^2$, for power-law exponent data in Panel a.

**Supplementary Discussion 3: Analytical expressions for active-inactive interface progression**

Our simple Fick's law-based model showed that the diffusion-like progression of the active-inactive interface was the consequence of diffusive dispersion of ATP (Fig. 3c). This result was based on numerical solutions of the Fick's law equation (Eq. 1); to gain deeper insight into the algebra underlying this modeling result, we considered the dynamics far from the boundary (i.e., discarding Eq. 2). The analytical solution to Fick's law (Eq. 1) subject to an initial step distribution is

$$C(x,t) = \frac{C_0}{2} \operatorname{erfc}\left(\frac{x - x_0}{2\sqrt{D\,t}}\right),$$  S5

where $x_0$ is the initial location of the step. We found the rate of progression of the interface by plugging Eq. S5 into Eq. 4 and solving for the active-inactive interface displacement $\Delta x$. Then the squared interface displacement can be written as:

$$\Delta x^2 = 4\,D\,t\,\left[\operatorname{erfc}^{-1}\left(\frac{2K}{C_0 + 2K}\right)\right]^2 = 2P_I\,t,$$  S6

where $\operatorname{erfc}^{-1}$ is the inverse of the complementary error function and the interface progression coefficient, $P_I$, is defined as

$$P_I \equiv 2D\,\left[\operatorname{erfc}^{-1}\left(\frac{2K}{C_0 + 2K}\right)\right]^2.$$  S7

Here, we found that, despite the nonlinearity of the flow speed-ATP conversion (Eq. 4), the interface progression coefficient depends linearly on the diffusion coefficient ($P_I \propto D$; Eq. S7) and the progression of the interface remains diffusion-like ($\Delta x^2 \propto t$; Eq. S6), which is consistent with our numerical results ($\gamma = 1$; Fig. 3c). Moreover, the derived interface progression coefficient $P_I(C_0)$ is almost identical to the numerical results (magenta curve and red dots in Fig. 3d). We thus reproduced the modeling results ($\gamma$ and $P_I$) algebraically.

**Supplementary Discussion 4: Effect of sample container height on the correlation lengths and times of flow velocities**

To characterize whether the mixing of our active fluid systems was driven by diffusion-dominated or convection-dominated active transport, we adopted a dimensionless quantity, the Péclet number: $Pe \equiv \bar{v} l_c / D$, where $D$ is the diffusion coefficient of ATP (Supplementary Discussion 1), $\bar{v}$ is the flow mean speed of active fluid, and $l_c$ is the correlation length of flow velocity of active fluid.[2,3] Determining $Pe$ requires $l_c$, and our previous work showed that increasing the sample container height could increase $l_c$.[14] Thus, we characterized how the correlation lengths and correlation times of active fluid flows depend on the sample container height.

To extract the correlation lengths in our sample, we prepared an active fluid sample with uniform activity, doped the sample with fluorescent tracer particles, and monitored and tracked these tracers to reveal the flow velocity fields of active fluid in the sample $\boldsymbol{V}(\boldsymbol{r}, t)$ as a function of time $t$ (Supplementary Fig. 5a). Then we calculated the velocity autocorrelation function,

$$\Psi(\Delta \boldsymbol{r}, \Delta t) \equiv \int d\boldsymbol{r} \, dt \, \boldsymbol{V}(\boldsymbol{r} + \Delta \boldsymbol{r}, t + \Delta t) \cdot \boldsymbol{V}(\boldsymbol{r} + \Delta \boldsymbol{r}, t + \Delta t), \tag{S8}$$

by deploying the convolution theorem

$$\Psi = \mathcal{F}^{-1}\{\mathcal{F}\{\boldsymbol{V}\} \cdot \mathcal{F}\{\boldsymbol{V}\}^*\}, \tag{S9}$$

where $\mathcal{F}\{\;\}$ represents the Fourier transform, $\mathcal{F}^{-1}\{\;\}$ represents the inverse Fourier transform, and $\Theta^*$ represents the complex conjugate of any variable $\Theta$. Then we normalized the correlation function as

$$\overline{\Psi}(\Delta \boldsymbol{r}, \Delta t) = \frac{\Psi(\Delta \boldsymbol{r}, \Delta t)}{\Psi(\boldsymbol{0}, 0)}, \tag{S10}$$

which allowed us to determine the normalized same-time velocity autocorrelation function as $\overline{\Psi}(\Delta \boldsymbol{r}, 0)$ (Supplementary Fig. 5b). To analyze the correlation lengths, we averaged the correlation function over orientations

$$\overline{\Psi}(\Delta r, 0) = \langle \overline{\Psi}(\Delta \boldsymbol{r}, 0) \rangle_{|\Delta \boldsymbol{r}| = \Delta r}, \tag{S11}$$

where $\langle \;\; \rangle_{|\Delta \boldsymbol{r}| = \Delta r}$ indicates averaging over the same magnitude of spatial displacement $\Delta \boldsymbol{r}$ (Supplementary Fig. 5c). Then we defined the correlation lengths $l_c$ as the separation distance where the normalized correlation function decayed to 0.5: $\overline{\Psi}(l_c, 0) \equiv 0.5$. Repeating the analysis of $l_c$ over the samples with heights varying from 60 to 700 μm revealed that the correlation lengths increased from 60 to 210 μm (Supplementary Fig. 5d). Given that the correlation lengths also represent the size of vortices in active fluid flow,[15] our analysis suggests that across our explored sample heights the vortices expanded by a factor of 3.5.

In addition to correlation lengths, our analysis also allowed us to extract the correlation time, $\tau_c$, which could reveal how rapidly the flow patterns changed (i.e., the lifetime of vortices). To determine the correlation time, $\tau_c$, we followed a similar analysis except we analyzed the normalized same-position temporal autocorrelation function $\overline{\Psi}(\boldsymbol{0}, \Delta t)$ (Supplementary Fig. 5c inset) and defined the correlation time $\tau_c$ as the time lapse when the normalized same-position temporal correlation function decayed to 0.5: $\overline{\Psi}(\boldsymbol{0}, \tau_c) \equiv 0.5$. Our analysis revealed that the correlation time remained ~20 seconds across our explored sample heights (Supplementary Fig. 5d inset), which suggests that the sample geometry did not play a significant role in the lifetime of the vortices. Overall, this work showed that increasing the sample container height enlarged the vortices but did not significantly affect their formation and deformation rates, which aligns with our previous studies about length scales of confined active fluid.[14]

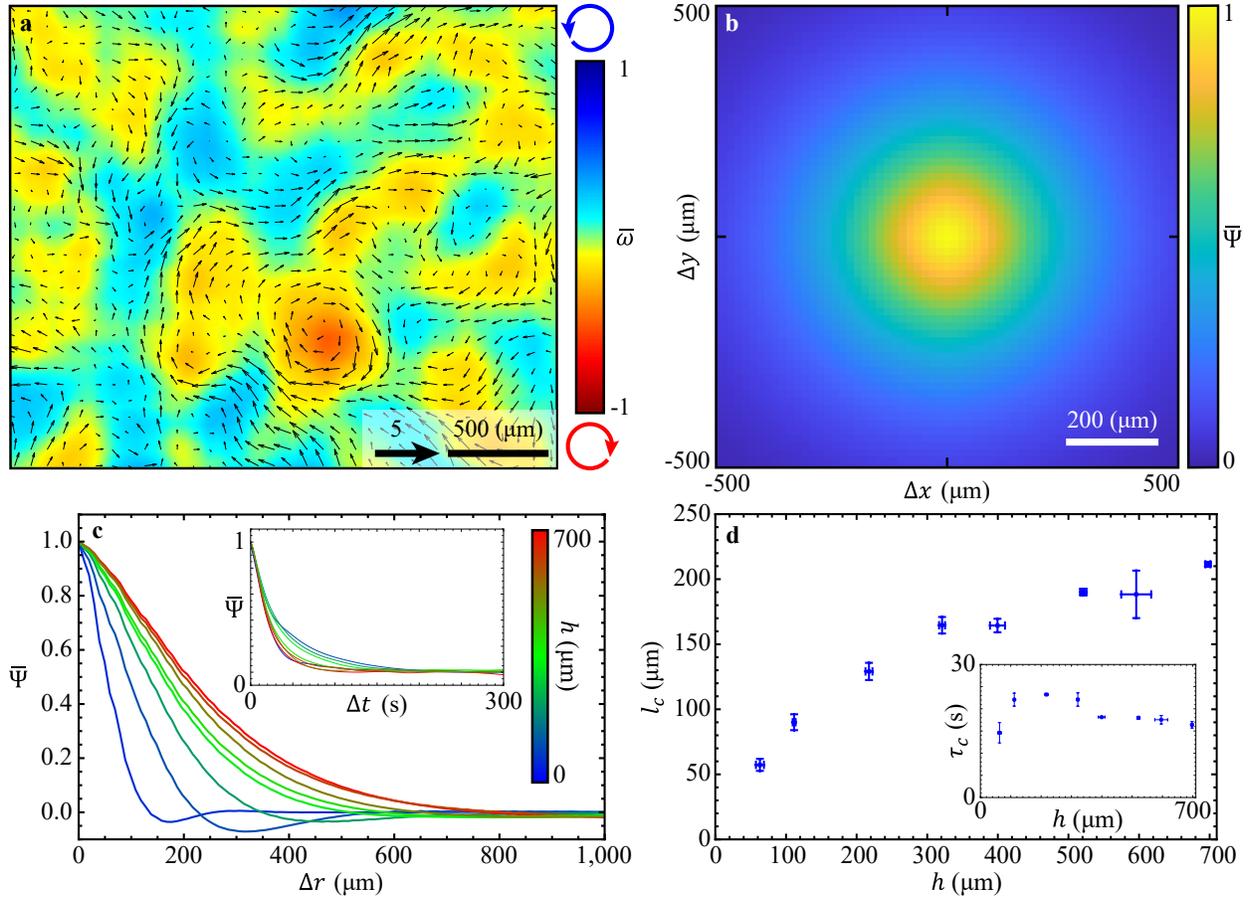

Supplementary Fig. 5: (Experimental results) Increasing sample container height increased the correlation lengths of flow velocities but did not significantly influence the correlation times. (a) Normalized velocity field and vorticity color map of active fluid flows in a 700-μm-thick flow cell. The velocity field $\boldsymbol{V}$ was normalized by the mean speed of active fluid flow; the vorticity $\omega \equiv [\boldsymbol{\nabla} \times \boldsymbol{V}]_z$ was normalized by the triple of standard deviation of vorticity, $\bar{\omega} \equiv \omega/[3 \, \text{std}(\omega)]$.[16] (b) Map of normalized same-time spatial autocorrelation of flow velocity in the same sample as Panel a. (c) Normalized same-time spatial autocorrelation functions of flow velocity as a function of separation distance (Eq. S11) for various sample heights. Inset: Normalized same-position temporal autocorrelation functions of flow velocity for various sample heights. (d) The correlation lengths of flow velocity increase with sample height. The error bars represent the standard deviations of two trials. Inset: The correlation times of flow velocity remained nearly invariant (~20 s) for sample heights from 60 to 700 μm.

**Supplementary Discussion 5: Mixing kinematics of activity-uniform active fluid**

This work focuses on mixing in active fluid systems with nonuniform distribution of activity. We showed that as the Péclet number of a system increases, the progression of the active-inactive interface changes from diffusion-like to superdiffusion-like (Fig. 4b) and the mixing times of suspended fluorescent dyes decrease (Fig. 5c). We decided that it would be elucidating to compare these results with those from active fluid systems with uniform activity distribution. A uniform active fluid system does not have an active-inactive interface, and thus we could not measure $\gamma$, but we measured mixing time of suspended fluorescein. We repeated the fluorescein-mixing experiments where the flow speed of active fluid were increased by increasing sample height (Fig. 5), but we performed the experiments in a system with uncaged ATP and thus flow speeds were uniformly distributed throughout the sample (Supplementary Fig. 6b inset) and analyzed the mixing time as a function of mean flow speed and Péclet number (Supplementary Fig. 6). Our analysis revealed that that the mixing time decreased with increasing Péclet number without a discernible transition, which we expected because a more convective active transport could mix the dye faster. Also, this result is similar to the one in active-inactive fluid systems (Fig. 5c), which suggested that the Péclet number was the controlling parameter for mixing time of suspended components in active fluid systems, regardless of the distribution of activity.

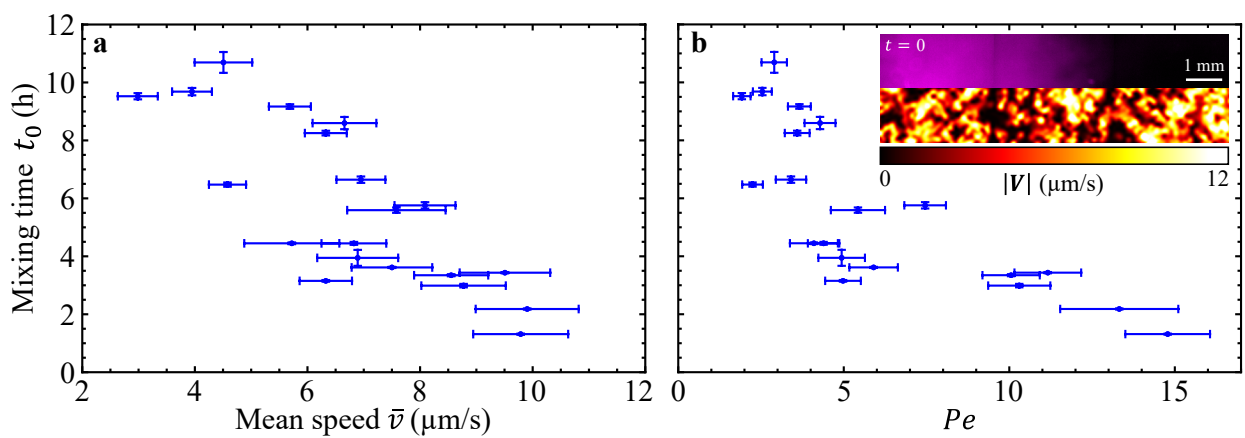

Supplementary Fig. 6: (Experimental results) Mixing time of UV-activated fluorescent dyes decreases with increasing Péclet number for activity-uniform active fluid systems. (a) Mixing as a function of mean speed of active fluid with uniform activity distribution. Accelerating active fluid flows accelerated the mixing process of suspended fluorescein, resulting in shorter mixing time $t_0$. Each dot represents one experimental measurement. Error bars in $t_0$ represent the slope fitting error as in Fig. 5b inset, and error bars in $\bar{v}$ represent the standard deviation of time-averaged flow speeds. (b) Mixing time in activity-uniform active fluid systems decreased with increasing Péclet number. The error bars are as in Fig. 5c. Inset: Micrograph of fluorescein uncaged by ultraviolet light exposure (magenta in upper half) and speed map of active fluid (lower half) in the beginning ($t = 0$) of the dye-mixing sample that had mean speed $\bar{v} \approx 9.8 \pm 0.8$ µm/s and mixing time $t_0 = 1.31 \pm 0.02$ h. Note that active fluid had uniform activity (flow speed distribution) in the beginning whereas the uncaged fluorescein was only distributed on one side (left) of the sample.

**Supplementary Discussion 6: Network melting mechanism may slow progression of active-inactive interface**

Our active-inactive fluid experiments showed that for the active-inactive interface to progress, not only did ATP need to be transported to the inactive fluid region, but also the inactive microtubule network needed to be activated from its inactive state (Supplementary Video 1). However, inactive microtubule networks behave like an elastic gel because the unfueled kinesin motor dimers are immobile and act as crosslinkers in the microtubule network,[8] and after ATP is transported to the active-inactive interface, it takes time for the fueled motors to fluidize or "melt" the inactive microtubule network[1] so the interface can progress. Thus, the active-inactive interface is expected to progress more slowly than it would if the network could melt instantly.

To examine this expectation, we analyzed the normalized speed profile expected on the basis of ATP distribution and compared them with the measured normalized speed profile. We assumed that the ATP was transported by the active fluid in the same way as fluorescein. Also, because ATP and fluorescein were both activated by UV light exposure, we assumed that the distribution of dyes was similar to that of activated ATP, which implied that the fluorescein and activated ATP had similar concentration profiles: $C_{ATP}(x,t)/C_{ATP0} \approx C_{fluorescein}(x,t)/C_{fluorescein0}$ where $C_X(x,t)$ represents the concentration of X and $C_{X0}$ represents the initial concentration of X. We also assumed that the gray values in the fluorescein micrographs were proportional to uncaged fluorescein concentration, which implied that the normalized concentration profile of uncaged fluorescein was similar to the normalized profile of gray values (G.V.) in the fluorescein micrographs: $C_{fluorescein}(x,t)/C_{fluorescein0} \approx \overline{G.V.}(x,t)$. Thus, we could deduce the profile of ATP from normalized profile of gray values in fluorescein micrographs: $C_{ATP}(x,t) \approx C_{ATP0}\overline{G.V.}(x,t)$. To extract the normalized profile of gray values in fluorescein micrographs, we considered the micrographs of fluorescein near the active-inactive interface (Supplementary Fig. 7a). Then we averaged the gray values of a micrograph vertically to get a profile of gray values, G.V.$(x,t)$ (Supplementary Fig. 7c). To normalize the gray value profile, we adopted the baseline model, which introduced two baselines as the upper and lower references of the gray values: $L_U(x,t)$ and $L_L(x,t)$, where $L_U$ was determined by fitting the gray value profile in the active bulk to a line (Supplementary Fig. 6c, magenta curve on top left) and $L_L$ was determined by fitting the gray value profile in the inactive bulk to a line (Supplementary Fig. 7c, magenta curve on bottom right). Then, the gray value profile could be normalized as

$$\overline{G.V.}(x,t) \equiv \frac{G.V.(x,t) - L_U(x,t)}{L_U(x,t) - L_L(x,t)}. \qquad S12$$

This normalization model has been commonly used in the analysis of thermal melting curves of DNA.[17] Here, we adopted this normalization model to reduce the influences of background light and nonuniform illumination in our profile analysis. Once the normalized gray value profile was determined, we could deduce the profile of ATP concentrations (Supplementary Fig. 7d). Then we converted the ATP concentration profile to flow speed profile of active fluid with the Michaelis-Menten equation (Eq. 4) followed by normalizing the flow speed profile (blue curve in Supplementary Fig. 7e). This is the profile deduced from distribution of uncaged fluorescein. To compare the deduced speed profile with the directly measured speed profile, we analyzed the tracer motion at the same time to extract the speed distribution of active fluid flow (Supplementary Fig. 7b), averaged the speed distribution vertically to extract the speed profile, and normalized the profile (red curve in Supplementary Fig. 7e). Our analysis showed that the normalized speed profile extracted from tracer motion (red curve in Supplementary Fig. 7e) fell behind the profile deduced from uncaged fluorescein distribution (blue curve), which demonstrated that the active-inactive interface progressed more slowly than expected from ATP distribution, and that the network melting played a role in the progression of the active-inactive interface and could slow down the progression.

The network melting mechanism was absent in our active-fluid hydrodynamic model; the model assumed that the network could melt almost instantly upon arrival of ATP (with negligible warm-up time from the initial development of activity-driven instability in the extensile $\boldsymbol{Q}$ field[1,18]), so we expected that the profile discrepancies observed in the experiment (Supplementary Fig. 7e) would not exist in our model. To examine the validity of our expectation, we analyzed the profiles both directly from flow speed distribution and as calculated on the basis of ATP distribution. In the simulation, we could directly access the ATP distribution (Supplementary Fig. 8a), which allowed us to determine the corresponding ATP concentration profile (Supplementary Fig. 8c inset). Then we converted the concentration profile to speed profile by the Michaelis-Menten equation (Eq. 4) followed by normalization to extract the normalized speed profile (blue curve in Supplementary Fig. 8c). To compare this ATP-based profile with the profile from flow speed distribution, we considered the speed map at the same time (Supplementary Fig. 8b), averaged the speed distribution vertically to get the speed profile, and normalizing the profile to get the normalized speed profile (red curve in Supplementary Fig. 8c). The modeling results showed that the ATP-based speed profile and the flow speed-based speed profile nearly overlapped across the active-inactive interface, which means that, in the simulation, ATP and activity progressed at the same pace. This is not consistent with experimental observations that the progression of activity fell behind ATP (Supplementary Fig. 7e). This mismatch between experiments and model results supported the existence of a network melting mechanism—in which the network needed to undergo a melting process before it could become fluidized—which was absent in the model.

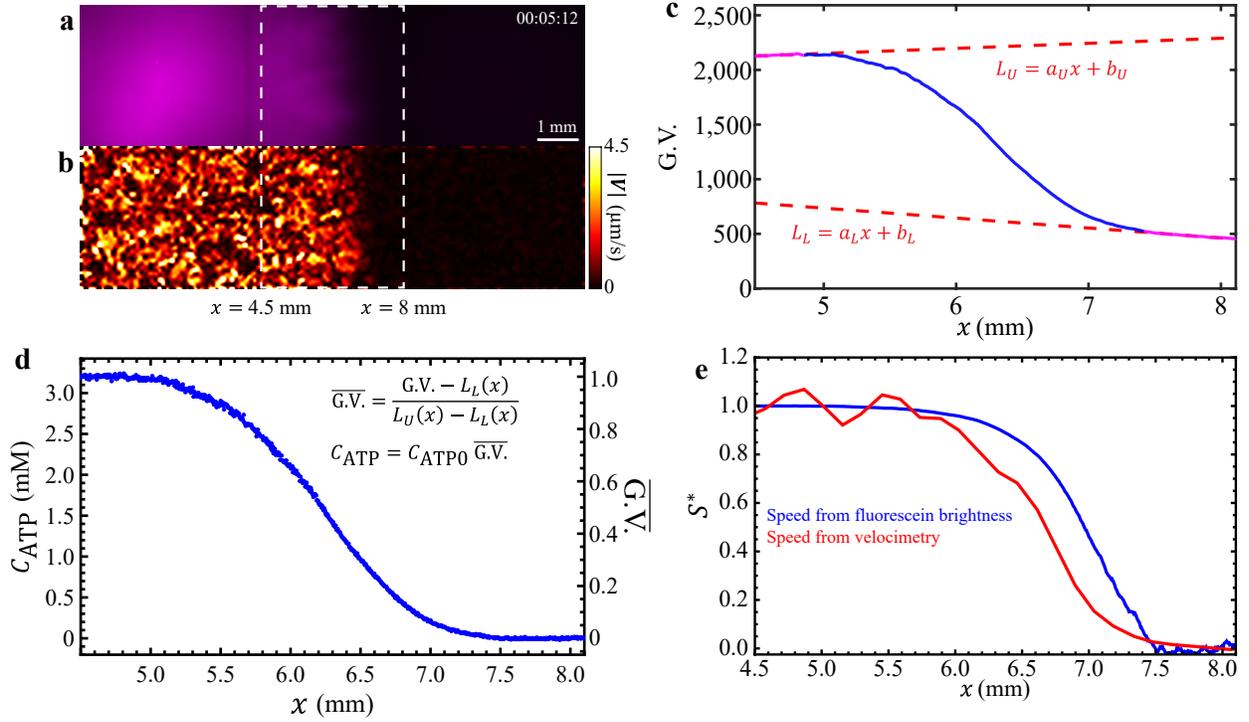

Supplementary Fig. 7: (Experimental results) The active-inactive interface progressed more slowly than expected from distribution of ATP. (a) Micrograph of fluorescein (magenta) in the active-inactive sample with an initial ATP concentration of $C_{ATP0}$ = 3.2 mM. The fluorescein was initially caged and thus did not fluoresce; exposure of the left side of the sample to ultraviolet light both activated the microtubule-kinesin fluid and uncaged the fluorescein, allowing it to fluoresce. The white dashed rectangle is the region of interest in the analysis in Panels c-e. The time stamp represents hour:minute:second. (b) The corresponding map of flow speed of active fluid deduced from tracking the motion of tracers. (c) Profile of gray values of the fluorescein micrograph in Panel a. The profile was normalized with a baseline model[17] that included an upper baseline determined by a line fitting to the gray value profile in the active bulk (the top-left magenta portion of the profile), $L_U = a_U + b_U$ where $a_U$ = 46 mm$^{-1}$ and $b_U$ = 1,920, and a lower baseline determined by a line fitting to the gray value profile in the inactive bulk (the bottom-right magenta portion of the profile), $L_L = a_L + b_L$ where $a_L$ = −91 mm$^{-1}$ and $b_L$ = 1,190. These two baselines served as the upper and lower references for profile normalization. (d) The gray values were normalized by the baseline model (right axis). The profile of ATP concentrations was deduced by scaling the normalized profile of gray values by the initial concentrations of ATP, $C_{ATP0}$ (left axis). (e) The profile of ATP concentrations was converted to the profile of flow speed by the Michaelis-Menten equation (Eq. 4). The speed profile was normalized (blue curve) as in Fig. 1e and the normalized profile was then compared with the profile measured from the velocimetry of tracers (red curve). The profile analyzed from the velocimetry fell behind the profile deduced from fluorescein brightness, which suggests that the melting mechanism of the crosslinked microtubule network slowed the progression of the active-inactive interface.

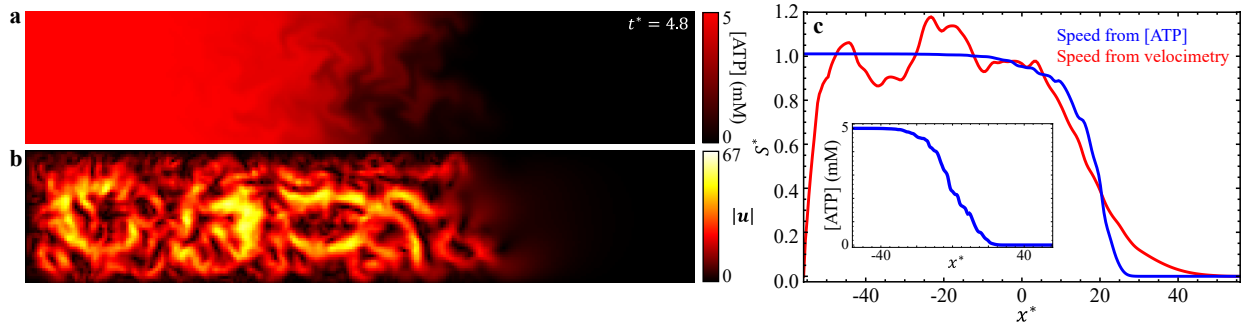

Supplementary Fig. 8: (Modeling results) In simulations of active-inactive fluid systems, profiles of active fluid flow and profiles deduced from ATP concentration overlapped. (a) Distribution of ATP concentrations in the active fluid simulation (Fig. 6) with $\alpha_0^* = 25$ and $D^* = 2$ at $t^* = 4.8$. (b) The corresponding distribution of flow speed. Note that the flow speed quickly dropped to zero (black) as it approached the boundaries because of the no-slip boundary condition. (c) The corresponding normalized speed profile analyzed from the distribution of flow speed (red curve) and the normalized speed profile deduced from the distribution of ATP (blue curve) nearly overlapped at the active-inactive interface. This suggests that, in the simulation, the active fluid was activated almost immediately after the ATP was transported to its location. Inset: The profile of ATP concentrations in Panel a, averaged vertically.

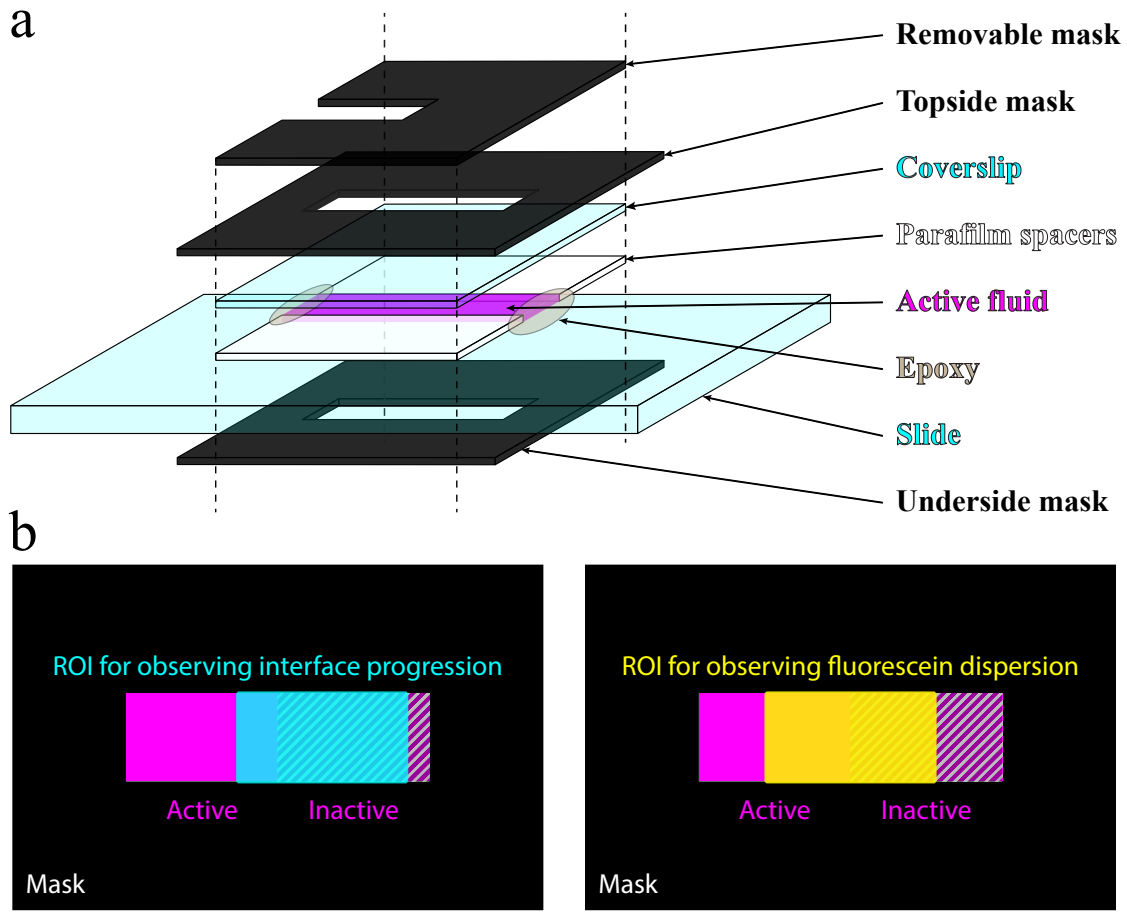

Supplementary Fig. 9: We used a removable mask to activate one side of the sample. (a) Active fluid was loaded into a glass flow cell consisting of a polyacrylamide-coated glass slide and coverslip with Parafilm as a spacer and sealed with epoxy. To prepare the active-inactive fluid system, we blocked one side of the sample (right half) with a removable mask. To prevent UV light from being scattered to the masked region by the epoxy and Parafilm, which could cause unwanted fluid activation in the masked region, we further blocked the rest of the sample, including epoxy and Parafilm, with 2 masks (one topside and one underside). After the UV exposure, we removed the mask to image the sample with fluorescent microscopy. (b) The sample is 20 mm long, which is wider than the field of view in our microscope even using a 4× objective, so we imaged 3 to 4 adjacent frames along the flow cell and stitched these frames into one large image. For the experiments monitoring active-inactive fluid interface progression (left), we selected the region of interest (ROI) as one quarter of active area and most of the inactive area (cyan rectangle) to observe the progression of the interface (Figs. 1–4). For the experiments monitoring the dispersion of fluorescein (right), we selected the ROI as half of the active and half of the inactive regions (yellow rectangle) to observe how one-sided dyes dispersed to the rest of the sample (Fig. 5).

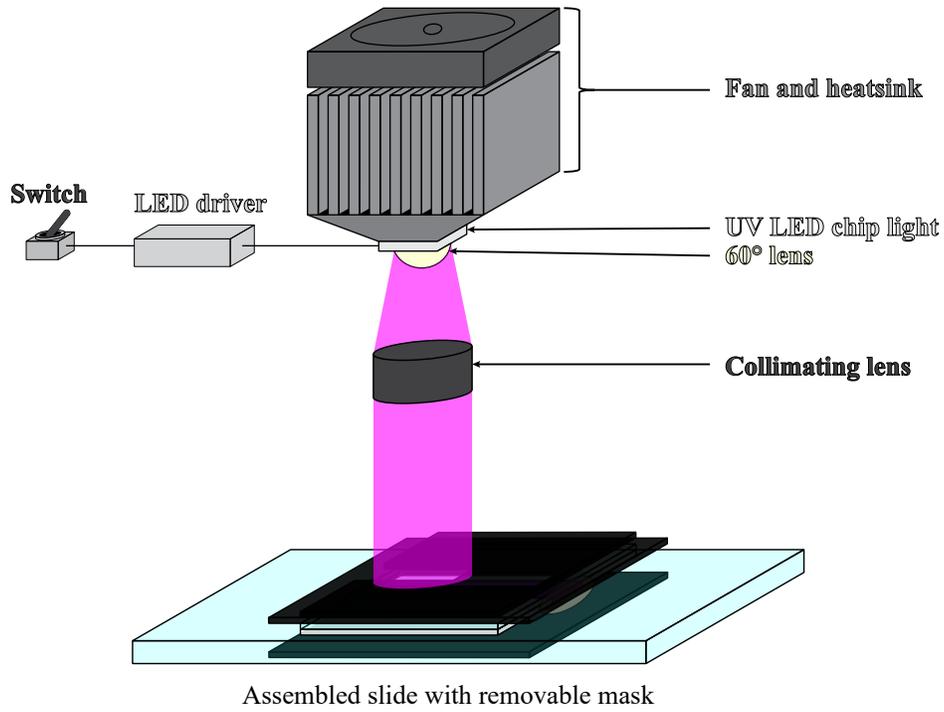

**Fan and heatsink**

UV LED chip light
60° lens

**Collimating lens**

Assembled slide with removable mask

Switch

LED driver

Supplementary Fig. 10: Setup to apply UV light to the masked sample (Supplementary Fig. 9a).[19] UV light was emitted with a UV LED chip (Amazon, B01DBZK2LM) powered by an LED driver (McMaster, 4305N124) and cooled with a fan-powered heatsink (Amazon, B01D1LD68C). To ensure that the light exposure was consistent across the sample, we parallelized the emitted UV light beams with a 60° lens (part of the heatsink) and a collimating lens (part of the microscope, Nikon, MEA54000).

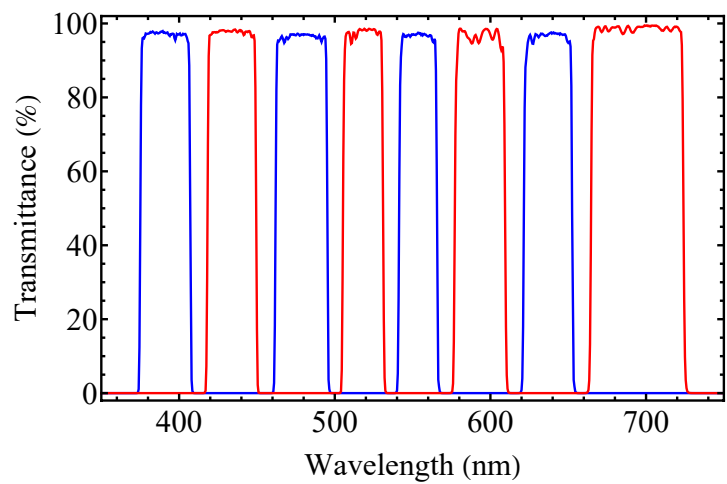

Supplementary Fig. 11: Excitation (blue) and emission (red) spectra of our multiband pass filter cube (Multi LED set, Chroma, 89402–ET). The filter cube is compatible with the fluorescent spectra of fluorescein (excitation: 490 nm; emission: 525 nm), Alexa 488 (excitation: 499 nm; emission: 520 nm), Alexa 647 (excitation: 650 nm; emission: 671 nm), and Flash Red (excitation: 660 nm; emission: 690 nm). Data source: www.chroma.com.

Supplementary Video 1: (Experimental results) Mixing of activated and inactive fluids. The fluid contained caged ATP, which could not fuel the kinesin motors until it was uncaged by exposure to ultraviolet light. After one side of the sample was exposed to ultraviolet light, the ATP molecules on that side of the sample were released and could fuel the kinesin motors to drive microtubules and create flows. The activated fluid blended with the inactive fluid until two fluids became one activity-uniform fluid. Cyan fibers are microtubules and red dots are tracers. The time stamp indicates hour:minute:second.

Supplementary Video 2: (Modeling results) Results of a one-dimensional, Fick's law-based model that simulates the mixing of active and inactive fluid under low Péclet number conditions ($Pe \lesssim 3$). The model describes how ATP distribution evolved from one side of a container to being uniformly distributed (top). The ATP was confined in a segment from $x = 0$ to $x = 20$ mm. The ATP distribution was converted to distribution of active fluid mean speed via Michaelis-Menten kinetics (Fig. 3b). The simulation shows that initially only one side of the system was activated, and then the system evolved toward an activity-uniform state (bottom). Active fluid with a higher initial concentration of ATP (8 mM; red curve) evolved toward an activity-uniform state faster than active fluid with a lower initial concentration of ATP (1 mM; black curve). The time stamp indicates hour:minute:second.

Supplementary Video 3: (Experimental results) Dispersion of UV-activated fluorescent dyes suspended in inactive (top) and active (bottom) microtubule-kinesin fluid. In the inactive system, the dyes were dispersed only by molecular diffusion, whereas in the active fluid system, the dyes were further transported by active fluid flows and thus dispersed through the sample more quickly. Time stamp indicates hour:minute:second.

Supplementary Video 4: (Modeling results) Simulated maps of ATP concentrations and flow speeds of active fluid for various pairs of dimensionless activity level, $\alpha_0^*$, and dimensionless molecular diffusion coefficient, $D^*$. In the no-activity system ($\alpha_0^* = 0$; top), dispersion of ATP was driven only by molecular diffusion ($D^* = 16$). When the fluid was activated ($\alpha_0^* = 25$; middle), the chaotic turbulence-like mixing flows were developed to actively transport ATP, which sped up the ATP dispersion. When the ATP diffusivity is increased ($D^* = 64$, bottom), dispersion of ATP was further accelerated. The simulation captured the roles of ATP diffusion and active fluid-induced convection in dispersing ATP.

Supplementary Video 5: (Experimental results) A checkerboard-pattern distribution of fluorescein and activity was achieved by applying UV light (00:00:12–00:01:06) in a checkerboard pattern to inactive fluid with caged ATP and caged fluorescein. The uncaged fluorescein (magenta in the left panel) was actively transported by flows driven by active microtubule network (cyan fibers in the middle panel) with the same checkerboard pattern of activity and reached a homogeneously-distributed state in 10 minutes (00:10:30). The right panel represents the merged images of fluorescein (left) and microtubules (middle). The grid size of the checkerboard was $a = 1$ mm. The time stamp indicates hour:minute:second.

Supplementary Video 6: (Modeling results) Simulated maps of ATP concentrations (top row) and flow speeds of fluids (bottom row) for active (right column) and inactive (left column) fluid systems where ATP was initially distributed in a checkerboard pattern with a dimensionless grid size of $a^* = 22$. The active fluid system ($\alpha_0^* = 25$; left) actively transported and homogenized ATP within the dimensionless time $t^* = 10$, while the ATP in the inactive fluid system ($\alpha_0^* = 0$; right), which relied on molecular diffusion ($D^* = 1$) to disperse ATP, did not reach the homogeneous state until the dimensionless time $t^* = 80$.